%% file: main.tex
\documentclass[twocolumn]{svjour3}

\usepackage{enumitem}
\usepackage{graphicx}
\usepackage{multirow}
\usepackage{multicol}
\usepackage{array}
\usepackage{makecell}
\usepackage{xcolor}
\usepackage{listings}
\usepackage{subfig}
\usepackage{graphicx}      
\usepackage{algorithm}
\usepackage{algpseudocode}
\usepackage{booktabs}
\usepackage{graphicx}
\usepackage{enumitem}
\usepackage[numbers]{natbib} 
\usepackage{amsmath}
\usepackage{amssymb}
\usepackage{hhline}
\usepackage{colortbl}
\usepackage{url}
\usepackage{hyperref}
\usepackage{xcolor}
\usepackage[english]{babel}

\newcommand{\eat}[1]{}

\newcommand{\ce}{CardEst}

\newcommand{\imdb}{IMDB}
\newcommand{\stats}{STATS}
\newcommand{\ergast}{ErgastF1}
\newcommand{\vg}{Genome}

\newcommand{\llm}{LLMs}

\newcommand{\reviewerone}[1]{\textcolor{black}{#1}} 
\newcommand{\reviewertwo}[1]{\textcolor{black}{#1}} 
\newcommand{\reviewerthree}[1]{\textcolor{black}{#1}} 
\newcommand{\reviewerfour}[1]{\textcolor{black}{#1}} 

\newcommand{\revision}[1]{{#1}}

\begin{document}
\titlerunning{Can Large Language Models be a Cardinality Estimator? An Empirical study} 
\authorrunning{Liangzu Liu et al.} 
\title{Can Large Language Models be a Cardinality Estimator? An Empirical study}

\author{Liangzu Liu$^{1,\dagger}$, Yiyan Wang$^{1,\dagger}$, Yinjun Wu$^{1,*}$, Runze Su$^1$, Zhuo Chang$^1$, Peizhi Wu$^{2 ,3}$, Jianjun Chen$^2$, Fuxin Jiang$^2$, Rui Shi$^{2}$, Bin Cui$^{1,*}$, Tieying Zhang$^{2,*}$ \thanks{$^{\dagger}$Liangzu Liu and Yiyan Wang contributed equally to this work.\\
$^{*}$Corresponding authors: Yinjun Wu, Bin Cui, and Tieying Zhang.\\
The source code of our proposed method is available at \url{https://github.com/PKU-SDS-lab/EA-LLM-IN-DB.git}.}}

\institute{$^1$Peking University, $^2$Bytedance, $^3$University of Pennsylvania}

\maketitle
\begin{abstract}
\input{abstract}
\end{abstract}

\input{intro}
\input{related_work}
\input{method}

\input{exp_setup}
\input{exp_benchmark}
\input{exp_generalization}
\input{exp_complex_queries}
\input{exp_training_efficiency}
\input{ablation}

\input{exp_end2end_time}
\input{discussion}
\input{conclusion}

\newpage
\bibliographystyle{plain} 
\bibliography{reference}

\end{document}

%% file: abstract.tex
Cardinality estimation (\ce) still remains a challenging problem for DBMS. Recent years have witnessed the success of ML-based cardinality estimators in outperforming traditional methods. However, these solutions suffer from poor generalizability to new data or query distribution, inability to handle complex queries, and substantial data preparation overhead, thus preventing their wide adoption in the real-world DBMS. Some recent efforts have been dedicated to addressing some but not all of these issues. 

We notice that the recent emerging Large Language Models (\llm) have shown their remarkable generalizability to unseen tasks, capabilities to understand complex programs, and power to perform data-efficient fine-tuning. In light of this, we propose to leverage \llm\ to mitigate the above issues. Specifically, we carefully craft prompts, and subsequently perform fine-tuning and self-correction during inference with \llm\ for CardEst task. We then extensively evaluate \llm' in-distribution and out-of-distribution generalizability, feasibility to support complex queries, and training data efficiency during fine-tuning \llm\ on pre-training datasets. The results suggest that \llm\ outperform the state-of-the-art in almost all settings, thus indicating their potential for the \ce\ task. We further measure the end-to-end query execution time in DBMS by using the estimated cardinalities of \llm\ in some practical settings, which suggests that the inference overhead of \llm\ can be outweighed by the benefits brought by \llm\ for \ce.

%% file: intro.tex
\section{Introduction}\label{sec: intro}
Cardinality estimation (\ce) is crucial for query optimization in DBMS \cite{garcia2008database, selinger1979access, leis2015good}. With more accurate cardinality estimation, DBMSs can generate more efficient query execution plans, thereby accelerating query processing. However, the default cardinality estimators in most commercial database systems such as Postgres \cite{postgresql2020} and MySQL \cite{mysql2020} employ simple statistics-based strategies \cite{poosala1997selectivity, selinger1979access, heimel2015self}, which, often struggle with suboptimal performance in practice. This is primarily due to their reliance on simplified assumptions, say, the independence between columns in one table, which fail to reflect the complexity of real-world databases.

To resolve this issue, some recent efforts have been dedicated to developing machine learning (ML) models to predict cardinalities. In comparison to non-ML-based solutions, these methods aim to either capture the sophisticated data distributions (namely data-driven methods) \cite{yang13deep, yang2020neurocard, hilprecht13deepdb, leis2017cardinality, park2020quicksel}
in databases or model the intricate correlations between a query and its cardinality directly (namely query-driven methods) \cite{kipflearned, woltmann2019cardinality}. Despite their remarkable success in reducing estimation errors, these solutions still suffer from the following challenges that prevent their wide adoption in practice.


First, most of these methods implicitly assume immutable database schema and static data distribution. Consequently, such models exhibit {\bf poor generalizability} to the scenario when the underlying {\bf data distributions are updated}, or when they are applied to \reviewerone{{\bf unseen workloads} ~\cite{negi2023robust,wu2024modeling, zeng2024price}} or {\bf unseen database instances}. As a result, to adapt these methods to these scenarios, one has to retrain data-driven models on the new data, or to prepare a completely new training set and then retrain query-driven models from scratch, which, is time-consuming. Although some data-driven methods such as \cite{wu2023factorjoin} can incrementally maintain their models to adapt to data updates, they still suffer from poor \ce\ performance in such scenarios as revealed by prior studies \cite{li2023alece} and our experimental results. 

\begin{table*}[!h]
    \small
    \centering
    \resizebox{\textwidth}{!}{
    \begin{tabular}{c|ccc|cc|c} \hline         \multirow{2}{*}{\ce\ method} & \multicolumn{3}{c|}{Generalizability} & \multicolumn{2}{c|}{Complex queries} & \multirow{2}{*}{
    \begin{tabular}{c}
Training data \\ efficiency
\end{tabular}
    } \\ 
         & data updates & workload shift & unseen database instances & LIKE predicates & GROUP-BY or DISTINCT queries &  \\ \hline
        {Data-driven 
        } & \checkmark\ (e.g., \cite{kim2024asm, hilprecht13deepdb}) & \checkmark & $\times$ & \checkmark\ (e.g., \cite{kim2024asm, wu2023factorjoin}) & $\times$ & High \\ \hline
        {Query-driven} & $\times$ & Partial (e.g., \cite{negi2023robust}) & $\times$ & \checkmark\ (e.g., \cite{negi2021flow}) & \checkmark\ (e.g., \cite{kipf2019estimating, hayek2020nn}) & Low \\ \hline
        {Hybrid} & {\checkmark\ (e.g., \cite{li2023alece})} & {\checkmark\ (e.g., \cite{zeng2024price})} & {\checkmark\ (e.g., \cite{zeng2024price})} & \checkmark\ (e.g., \cite{aytimur2024lplm}) & $\times$ & Low \\ \hline
       \llm\ & \checkmark & \checkmark & \checkmark & \checkmark & \checkmark & High \\ \hline
    \end{tabular}
    }
    \caption{Characteristics of state-of-the-art ML-based cardinality estimators and our method}
    \label{tab:summarize_sota}
\end{table*}


In addition, the majority of the mainstream methods are {\bf limited to estimating cardinalities for simple queries, i.e., Selection-Projection-Join (SPJ) queries with selection predicates only for categorical and numeric attributes}. This class of queries is referred to as {\em simple SPJ queries} throughout this paper. In contrast, the need to estimate cardinalities for complex queries is quite common in practice \cite{lee2023analyzing,krishnan1996estimating,chaudhuri2004selectivity,aytimur2018estimating,jagadish1999substring}. 
\reviewerone{Data-driven methods, such as \cite{hilprecht13deepdb,yang2020neurocard,kim2024asm} struggle with complex queries, e.g., GROUP-BY queries, in particular when distinct value counts are large, making them inefficient. Query-driven methods, while more flexible, require extensive feature engineering and operator-specific mechanisms to achieve acceptable performance~\cite{kipf2019estimating, aytimur2024lplm}.}
Furthermore, {\bf the data preparation process is cumbersome} for state-of-the-art ML-based cardinality estimators, in particular, for query-driven methods. 
One essential step for such methods is to execute an \reviewerfour{large} amount of queries to collect their ground-truth cardinalities as the training labels, which could be extremely slow. Note that although data-driven methods don't rely on any queries as input, they still severely suffer from the above two issues.


To overcome the above challenges, some recent hybrid solutions \cite{negi2023robust,zeng2024price,kurmanji2023detect} were proposed by incorporating both data distribution and query information. 
However, they are still limited to SPJ queries rather than general SQL queries. Additionally, their 
training (or pre-training) phase is data-hungry. 
For example, up to 1.3$\times$10$^6$ SQL queries are needed for pre-training in \cite{zeng2024price}.
On the other hand, some recent hybrid and query-driven works strive to estimate cardinalities for more expressive queries, such as GROUP-BY queries \cite{kipf2019estimating} or DISTINCT queries \cite{hayek2020nn}\footnote{Note that there is no difference between DISTINCT queries and GROUP-BY queries regarding their cardinalities} or queries with LIKE predicates \cite{aytimur2024lplm, negi2021flow}. However, these are ad-hoc solutions for those specific operations, thus incompatible with other techniques to support general SPJ queries.
We summarize the characteristics of such representative works in Table \ref{tab:summarize_sota}. As suggested by this table, the aforementioned issues in \ce\ task cannot be resolved simultaneously in any of the existing ML-based methods, thus hindering their broad usage in the real-world DBMS.


In this paper, we propose to tailor Large Language Models (\llm) to the \ce\ task. This is motivated by their outstanding generalizability \cite{mosbach2023few}, capability of understanding expressive programs \cite{chen2021evaluating}, 
and power of performing data-efficient fine-tuning \cite{mosbach2023few}.
To our knowledge, this is the first paper to investigate the feasibility of leveraging \llm\ for \ce. 
However, this is not straightforward.
To accomplish this goal, we designed a series of preliminary strategies to adapt \llm\ to \ce\ task.

First,  
to incorporate sufficient information about the underlying data distribution into \llm\ while respecting their context window limits, 
we propose to only include coarse-grained statistics from databases along with cardinality estimates from other \ce\ methods in the prompts.  
Second, considering that \llm\ may produce arbitrary tokens beyond numeric ones, we propose to perform fine-tuning on \llm\ to increase their likelihood of generating correct digit numbers. Plus, considering that \llm\ may output severely wrong cardinalities or even non-numeric tokens, namely \emph{hallucination} \cite{rawte2023troubling,maynez2020faithfulness}, during the inference phase, we thus further propose to leverage the self-correction mechanism \cite{madaan2024self} to mitigate this issue.

Building on top of the proposed solution, we then perform a comprehensive empirical study to assess the capability of \llm\ as a cardinality estimator in various settings. In particular, we first evaluate the in-distribution generalizability of \llm\ on estimating cardinalities for SPJ queries on standard benchmark datasets (Section \ref{sec: benchmark}), and then the out-of-distribution generalizability of \llm\ to novel query patterns, new data distributions and unseen database instances in Section \ref{sec: generalizability}. Subsequently, we examine the capabilities of \llm\ in performing \ce\ for complex queries such as LIKE queries and DISTINCT queries in Section \ref{sec: complex_queries}. Furthermore, we vary the number of training samples used for fine-tuning \llm\ for \ce\ in Section \ref{sec: data_efficiency}, which aims to investigate the data efficiency of \llm. The effectiveness of different components of our preliminary solutions is studied in Section \ref{sec: ablation}. Overall, the results in the above experiments suggest that \llm\ outperform baseline methods almost all the time, thus demonstrating {\bf \llm' in-distribution and out-of-distribution generalizability, feasibility to support complex queries, and superior data efficiency in fine-tuning}. 

We then conclude the empirical studies by evaluating the end-to-end query execution time of the query plans produced by using the cardinalities estimated by \llm, as detailed in Section \ref{sec: end_to_end}. 
Considering the significant inference overhead of \llm, we further propose to leverage \llm\ to only estimate cardinalities for those potentially costly queries rather than all of them within a workload. 
Our findings in Section \ref{sec: end_to_end} suggest that this strategy achieves shorter end-to-end running time compared to baseline methods even when accounting for the model inference overhead. 
These results underscore the potential of integrating \llm\ into real-world DBMS. 
We summarize our contributions as follows:
\begin{itemize}[noitemsep, topsep=0pt, left=0pt]
    \item We attempt to adapt \llm\ to \ce\ task. To accomplish this, we carefully craft prompts and perform fine-tuning and self-correction during inference time with \llm.
    \item We extensively evaluate the generalizability, the capability for supporting complex queries and training data efficiency of \llm\ to demonstrate their effectiveness for \ce\ task. 
    \item We propose a practical strategy to employ \llm\ for end-to-end query execution, which justifies their potential to be a practical \ce\ method.
\end{itemize}

\reviewerfour{Our method can be categorized as a hybrid \ce\ method, as it relies on query inputs, requires pre-fine-tuning and incorporates data distribution information in prompts. 
This design bridges the advantages of both paradigms and enables better generalization across diverse query types.}


%% file: related_work.tex
\section{Related work}\label{sec: related_work}

\paragraph{\ce\ for simple (SPJ) queries\reviewerfour{.}} Cardinality estimation is a long-standing problem for query optimization. Some recent efforts have been committed to leveraging machine learning models to enhance the \ce\ performance, primarily for Selection-Projection-Join (SPJ) queries with selection predicates only for categorical and numeric attributes. These methods are generally categorized into three groups, i.e., Data-driven \cite{yang13deep, yang2020neurocard, hilprecht13deepdb, leis2017cardinality, park2020quicksel}, Query-driven \cite{kipflearned, woltmann2019cardinality}, and hybrid methods \cite{li2023alece, zeng2024price}. The data-driven methods aim to model the correlations between columns in the database with unsupervised models, which is followed by estimating the probability of tuples occurring in the query result. On the other hand, query-driven methods target building supervised models to directly predict the cardinalities of input queries, which 
can be further extended to hybrid methods \cite{li2023alece, zeng2024price} by constructing additional modules to explicitly model data distribution.
\reviewerone{However, data-driven methods, such as those based on sum-product networks \cite{hilprecht13deepdb} and deep autoregressive models \cite{yang2020neurocard,kim2024asm}, struggle to process GROUP-BY queries due to inherent design limitations. Specifically, they could be highly inefficient when the distinct value count of individual columns or combinations of columns is large. 
On the other hand, query-driven ones can be more flexible and efficient in handling these complex queries. However, they also have practical limitations. In particular, they rely on operator-specific mechanisms for encoding operators and collecting relevant distribution-aware statistics. This requires extensive feature engineering 
to reach an acceptable performance~\cite{kipf2019estimating, aytimur2024lplm}.}

\paragraph{\ce\ for complex queries\reviewerfour{.}}
Some \ce\ methods have been proposed for handling complex queries \cite{lee2023analyzing,krishnan1996estimating,chaudhuri2004selectivity,aytimur2018estimating,jagadish1999substring}. For instance, 
\cite{kipf2019estimating} attempts to adapt the query embedding module of \cite{kipflearned}, one representative query-driven method for simple queries, to support GROUP-BY queries. 
\cite{kwon2022cardinality,aytimur2024lplm,shetiya2020astrid} \reviewerfour{aim to estimate} cardinalities for queries with LIKE predicates.
In addition, \cite{hayek2020nn} proposed initial solutions to deal with queries with OR, DISTINCT and NOT operators. However, these solutions still suffer from generalizability and data inefficiency issues. Moreover, these ad-hoc solutions might be incompatible with other CardEst models of simple queries.


\paragraph{\ce\ under changing workloads or data distributions\reviewerfour{.}}
Most state-of-the-art ML-based cardinality estimators struggle with the dynamically changing data and query distribution \cite{wang2021we}, namely data distribution shift and workload shift. To mitigate the negative effect of workload shift, \cite{li2022warper,wu2024modeling} proposed to perform further fine-tuning or retraining with carefully selected queries, while \cite{negi2023robust} proposed to emulate the workload shift during the model training phase so that the learned model can tolerate some amount of workload shift. \reviewerone{Additionally, \cite{negi2023robust} extends MSCN \cite{kipf2019estimating}, a classical query-driven method to handle very slight workload shifts, and ALECE \cite{li2023alece} aims to train a generalizable model to deal with query workloads involving data update operations. }

On the other hand, to deal with data distribution shift, one can build \ce\ models that are generalizable to updated data distributions. For instance, \cite{li2023alece} designed an attention-based model to embed the underlying data, which can accommodate the data update operations in the workload. \cite{zeng2024price} builds a pre-trained model that adapts to new database schemas. \reviewerone{DDUps~\cite{kurmanji2023detect} seeks to alleviate the issue of data shifts for data-driven approaches by transfer learning.} \reviewerthree{Another line of strategies to deal with data shift is to employ solutions such as \cite{hilprecht13deepdb, 2021iris} to incrementally update \ce\ models to reflect the updates of data distributions. However, \cite{hilprecht13deepdb} is highly ineffective since they have to perform updates tuple-by-tuple while \cite{2021iris} can only support a very limited type of queries.}

\reviewerone{
Some recent hybrid solutions have been proposed to deal with workload and data distribution shift simultaneously. For instance, PRICE \cite{zeng2024price} builds a pre-trained \ce\ model for achieving generalizability to unseen query templates and database schemas. } \reviewerthree{In addition, \cite{wehrstein2025towards} proposed a foundation model for databases, which aims to learn generalizable representations of data and query plans such that they can be easily adapted to a variety of downstream database tasks. 
However, all of these solutions are limited to simple SPJ queries. 
}

\paragraph{\llm\ for other database tasks\reviewerfour{.}} Due to the exceptional power of \llm, they have been adopted for various database tasks, including knob tuning \cite{li2024large,huang2024llmtune}, database diagnosis
\cite{singh2024panda, zhou2023llm, zhou2024d}, query rewrite \cite{li2024llm,liu2024query,zhou2024db} and text-to-SQL \cite{li2024codes, fan2024combining, trummer2022bert}. Most of them aim to leverage the few-shot or zero-shot reasoning capability of off-the-shelf \llm\ without any fine-tuning. In contrast, we discover that fine-tuning \llm\ is essential for guaranteeing \ce\ performance. 

%% file: method.tex
\section{Cardinality Estimation with \llm}\label{sec: method}

\subsection{Problem definition}
\begin{definition}{\bf (Cardinality Estimation (\ce))}
Given a database $D$, and a query $Q$, our problem is to estimate the cardinality Card(Q) of Q, i.e., the result of SQL ``select count(*) from $Q$''.
\end{definition}
Note that, unlike prior experimental studies on cardinality estimation (\ce), e.g., \cite{kim2022learned, han2021cardinality} which only focus on select-project-join (SPJ) queries, we target 
solving \ce\ problem for arbitrary classes of queries, e.g., queries with DISTINCT operations or LIKE predicates (referred to as DISTINCT queries and LIKE queries hereafter). Plus, ideal \ce\ methods, in particular, ML-based ones, should be generalizable to unseen query patterns, data updates, and unseen datasets. Finally, building \ce\ models should be data-efficient, meaning training such models should not rely on large amounts of data that require extensive query generation time.

\subsection{Overview of tailoring \llm\ to \ce\ task}\label{sec: method_overview}
We first present the overview of the preliminary solution of tailoring \llm\ to \ce\ task, in particular, how to \emph{prompt, fine-tune, and \reviewerfour{perform} inference} with \llm. The details of these components are provided in Section \ref{sec: glm_prompt}-\ref{sec: glm_inference}. In this subsection, we highlight some critical challenges of using \llm\ for \ce\ task and briefly discuss how they can be addressed with the proposed solution. 

First of all, we craft appropriate prompts for \ce, which targets incorporating all the information needed for cardinality estimation. Besides the input queries, recent studies also suggest the 
necessity of incorporating data information that 
captures complex correlations across columns or even across joinable tables for accurate cardinality estimation \cite{kipflearned, li2023alece, hilprecht13deepdb}. 
However, the amount of such data information grows substantially with the increasing sizes of database instances and numbers of dimensions/columns, which is difficult to fit within the {\em limited context windows} of \llm\ (e.g., GPT-4o only accepts up to 32K tokens \cite{openai2023gpt4o} in the prompt). 
To mitigate this issue, we propose plugging cardinality estimates from other \ce\ models into the prompt.
Cardinality estimates implicitly carry basic information about data distribution. 
Interestingly, this can be also viewed as being inspired by \cite{xu2023small} which {\bf integrates small model predictions for enhancing their performance}.  \reviewertwo{Note that this shares the same spirit as the methods proposed by previous studies, such as \cite{wu2020bayescard,liu2021fauce}, which primarily aim for robust cardinality estimations by incorporating the output of multiple cardinality estimators}. We further propose a confidence-based strategy to address the issue of possibly unreliable estimates produced by other \ce\ models.

Moreover, considering that {\em \llm\ may generate arbitrary text tokens while cardinalities are numeric values}, we thus propose to {\bf fine-tune \llm} to guide them to only output numeric tokens as estimated cardinalities. Specifically, instead of predicting cardinalities as numbers, we regard cardinality numbers as numeric token sequences and aim to maximize the likelihood of those sequences.

However, the fine-tuned \llm\ may be still error-prone, say producing an unreasonable cardinality number
, and even generating non-numeric tokens, during the inference phase. These issues can highly likely occur especially when {\it out-of-distribution query workloads or data distributions} occur, thereby diminishing \ce\ performance. To mitigate this issue, whenever such prediction errors happen, we leverage the {\bf self-correction mechanism} discovered by \cite{madaan2024self} to iteratively guide \llm\ to refine their output such that the output tokens can compose cardinality numbers that are not far away from the predicted cardinalities by other estimators.


\subsection{Prompting \llm\ for \ce}\label{sec: glm_prompt}

\begin{figure}
    \centering
        \includegraphics[width=0.9\linewidth]{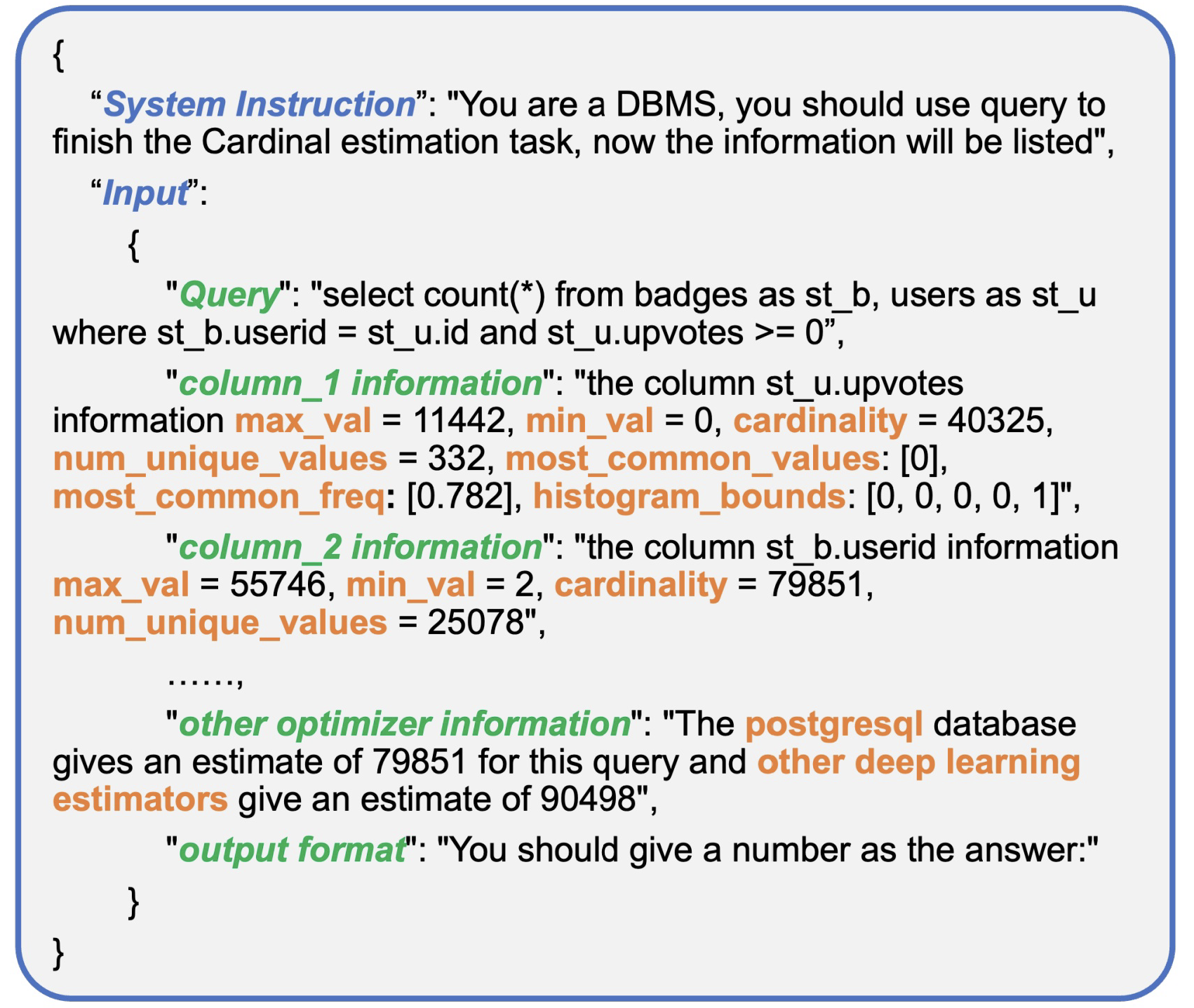}
    \caption{One example of prompts for performing the cardinality estimation task with \llm}
    \label{fig:prompt}
\end{figure}

As mentioned above, we first construct prompts to attempt to leverage both the query and data information for estimating the cardinality of one query. The crafted prompts are outlined in Figure \ref{fig:prompt} with one example,  which contains the following elements:
\begin{itemize}[noitemsep, topsep=0pt, left=0pt]
    \item ``System instruction'' describes the role of \llm\ as a cardinality estimator. 
    \item ``Input'' contains the input query (denoted by the ``Query'' field) and other essential information that is essential for the reasoning process of \llm, including:
    \begin{itemize}[noitemsep, topsep=0pt, left=0pt]
        \item Coarse-grained single-column statistics (denoted by ``column\_k'' field with k=1,2,...), e.g., the maximum, the minimum, the number of unique values, most frequent values and their frequency of each column appearing in the input query (represented by ``histogram\_bounds'' in Figure \ref{fig:prompt});
        \item Estimated cardinalities by other \ce\ methods (denoted by ``other estimator information''), including the ones from the default cardinality estimator within Postgres database \cite{PostgreSQL1996} and other ML-based \ce\ models;
        
    \end{itemize}
\end{itemize}

\begin{figure}[!h]
    \centering
    \includegraphics[width=1\linewidth]{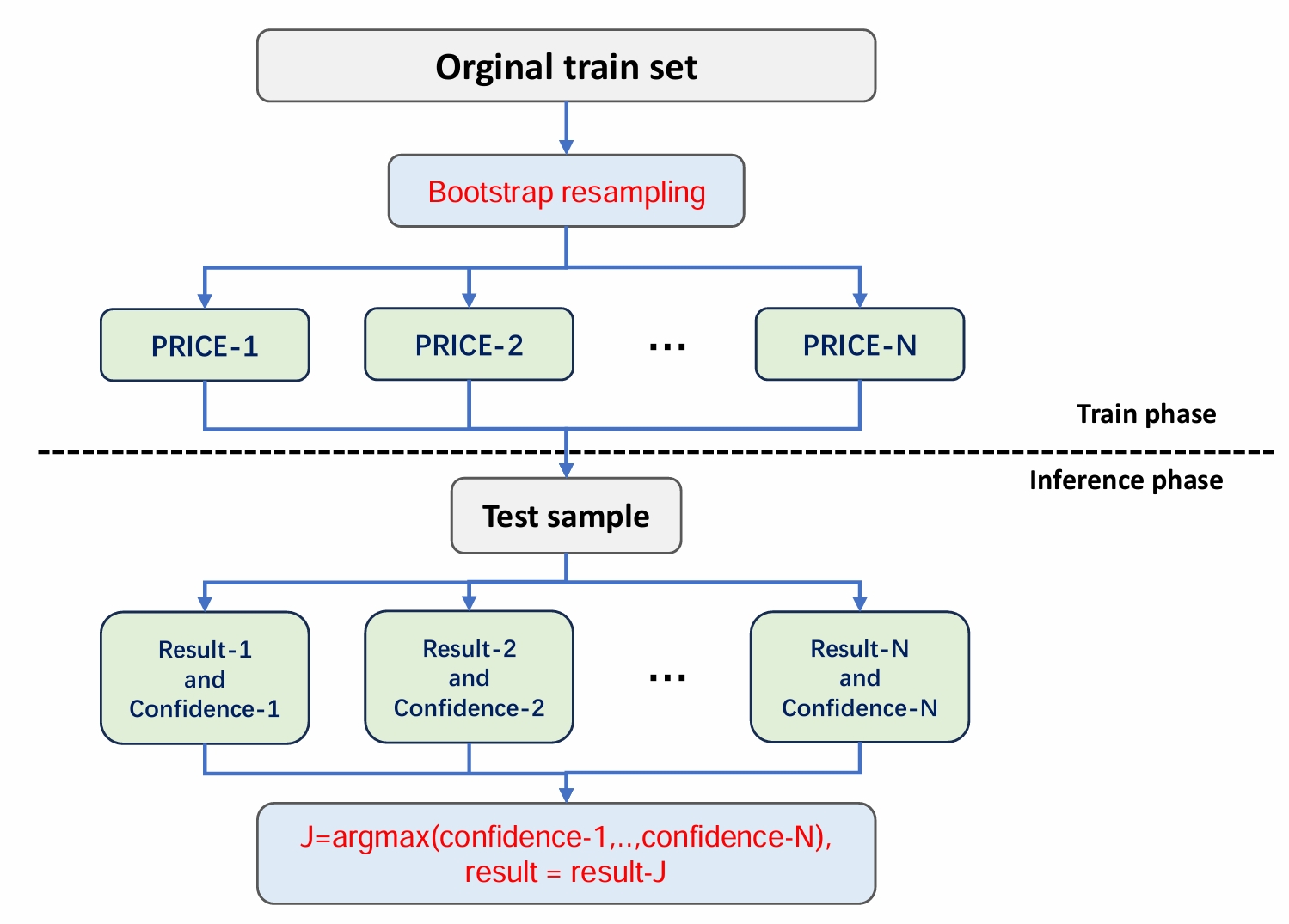}
    \caption{Other estimator results selection methods}
    \label{fig:PRICE-prompt}
\end{figure}

\reviewertwo{For string columns, instead of constructing histograms, we adopt simple coarse-grained statistics such as the number of unique values and the most frequent strings with their corresponding frequencies, which provide sufficient information for handling LIKE queries effectively.}
Note that SOTA ML-based cardinality estimators usually require complicated statistics from the underlying databases as input, such as the complete histogram of each column \cite{li2023alece, zeng2024price, kipflearned}
or even the fanout factor of each value $v$ in a joinable column, 
However, due to the limited context window size, it is impossible to include all of such detailed statistics in the prompts. But considering that such statistics are proven beneficial for enhancing \ce\ performance
\cite{zeng2024price}, we thus included the estimated cardinality values provided by the state-of-the-art models in the prompts as described above. 
Since the correctness of the estimated cardinalities heavily depends on how the underlying data distribution is captured by the models, this can be regarded as one indirect way of incorporating complicated statistics into prompts. Also, this can be viewed as the integration of small model predictions into \llm\ for performance enhancement as introduced in \cite{xu2023small}. But note that such estimated cardinalities from other \ce\ methods may be inaccurate, thus potentially hurting the \ce\ performance of our method. Hence, it would be ideal if such estimated cardinalities were reliable enough for each input query, which could be the ones with high prediction confidence \cite{angelopoulos2023prediction}. To accomplish this, we train multiple versions of the PRICE models, one SOTA model for \ce~\cite{zeng2024price}, such that at least one PRICE model produces enough confident estimates for {\em every input query}. 
As illustrated in Figure \ref{fig:PRICE-prompt}, we performed bootstrap resampling on the original training set to train multiple versions of the model, named PRICE-1, PRICE-2, \dots, PRICE-N. By applying these models to each test sample, we obtain the estimated results (result-1, result-2, \dots, result-N), and the confidence scores (confidence-1, confidence-2, \dots, confidence-N).

\revision{These confidence scores are quantified by the size of the confidence intervals produced by bootstrap resampling \cite{dixon2006bootstrap} for the estimated cardinalities for each query using each version of the model. Formally, the confidence score $confidence-i$ for model PRICE-$i$ is defined as the inverse of its prediction uncertainty for each query:
$${confidence-i} = \frac{1}{CI_{high,i} - CI_{low,i} + \epsilon}$$
The above confidence interval $[CI_{\text{low},i}, CI_{\text{high},i}]$ as follows. First, we repeatedly estimate the cardinality using multiple independent runs on the same model PRICE-i with different random seeds. Then, using bootstrapping resampling on these estimates, we construct an empirical distribution. The lower and upper bounds, $CI_{\text{low},i}$ and $CI_{\text{high},i}$, are taken as the $2.5^{\text{th}}$ and $97.5^{\text{th}}$ percentiles of this bootstrap distribution, corresponding to a $95\%$ confidence level. The interval width is defined as $CI_{\text{high},i} - CI_{\text{low},i}$. A small constant $\epsilon$ is added to the width where necessary to maintain numerical stability.
}

Generally speaking, smaller confidence intervals indicate higher confidence scores.
We then identify the most confident results by calculating $$J = \arg\max(\text{confidence-1}, \dots, \text{confidence-N}),$$ which is then employed to determine the final estimated result used for the prompt, i.e., \( \text{result} = \text{result-}J \). \reviewertwo{We adopt bootstrap resampling in PRICE following its standard statistical purpose—to quantify estimation uncertainty and improve robustness rather than to aggregate multiple weak predictions. By generating multiple estimates from sampled data and selecting the one with the narrowest confidence interval, the final output is statistically stable, less sensitive to data noise, and less prone to overfitting, without altering the underlying model strength.}

\reviewerthree{However, it is important to note that PRICE is not applicable for all settings. Whenever such scenarios occur, we either rely solely on the estimated cardinalities provided by Postgres or employ the estimates from the best baseline methods. This approach yields satisfactory \ce\ performance, as demonstrated in Section \ref{sec: gen_data_dist} and Section \ref{sec: complex_queries}.}

We also attempt to incorporate training queries that resemble the input query in the prompt, which are regarded as few-shot examples for \llm. 
However, as revealed in Section \ref{sec: ablation}, including these few-shot examples may hurt \ce\ performance, which is thus disabled by default.

\begin{figure}[!h]
    \centering
    \includegraphics[width=0.8\linewidth]{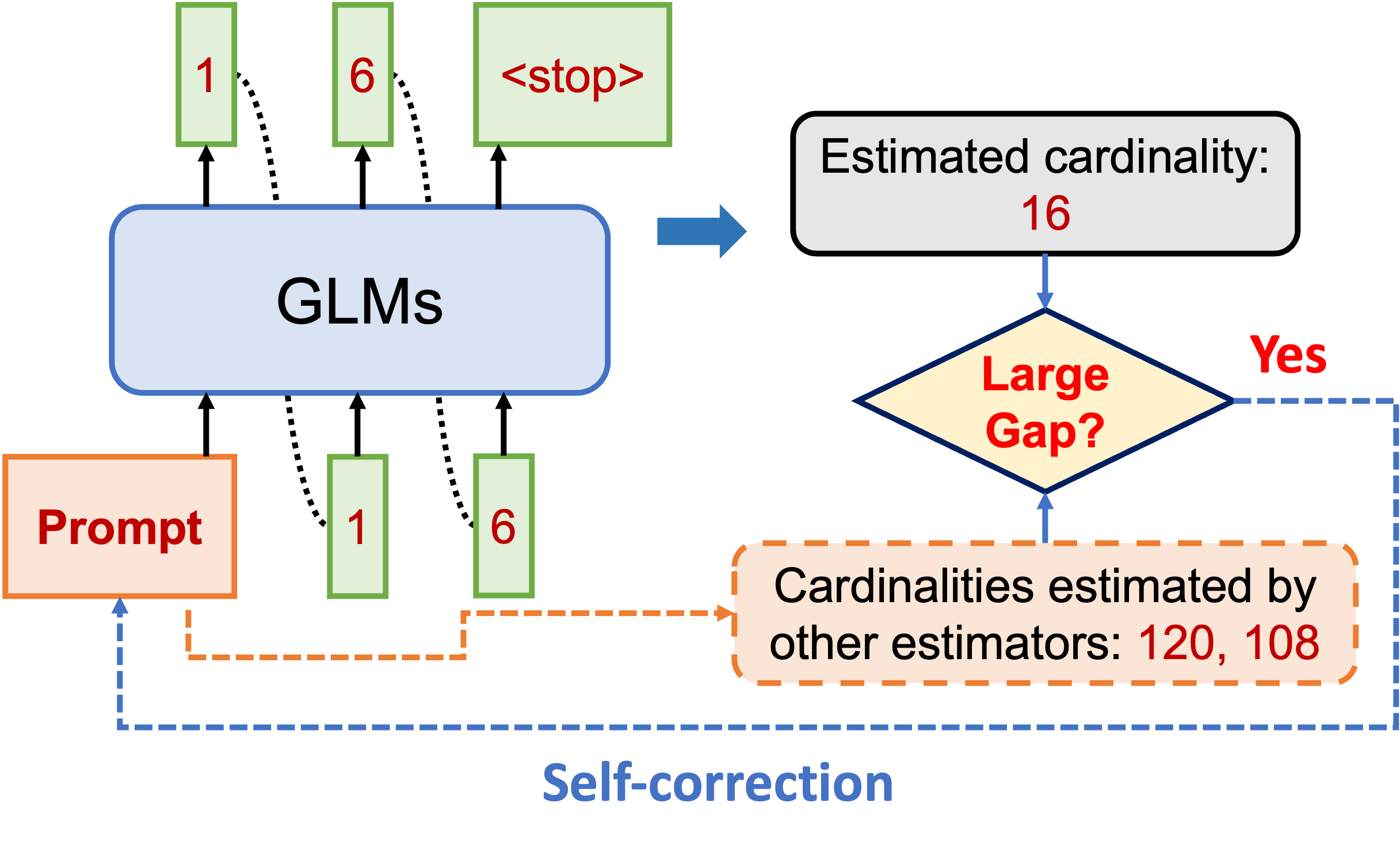}
    \caption{Inference process of \llm\ for \ce}
    \label{fig:inference}
\end{figure}

Given the above prompts, we anticipate that \llm\ would output a numeric token sequence as the estimated cardinality, which is performed in an autoregressive manner. This process is visualized in Figure \ref{fig:inference} with one example (ignore the ``self-correction'' for now). In this figure, \llm\ take the above prompts as input and output one digit ``1'' first, which is then appended to the prompt as the input to \llm\ for predicting the next digit ``6''. This process is repeated until the terminating token ``$<$stop$>$'' appears. All the generated digits are then combined as the estimated cardinality, ``16''.

\reviewerone{We adopt digit-wise generation to ensure consistent tokenization and generalization across arbitrary numbers.} \reviewerfour{ Multi-digit tokens like “16” may not exist uniformly in LLM vocabularies, leading to inconsistent tokenization. } \reviewerone{Generating numbers one digit at a time leverages the autoregressive nature of \llm\, allowing them to compose any integer from basic numeric tokens containing numbers 0 to 9. In contrast, if we force an LLM to output a whole number as a single token, this would require adding a new token and embedding for every possible integer, which is not only impractical to train but also impossible to generalize.} \reviewerfour{During fine-tuning, the model is restricted to output single digits followed by a ``$<$stop$>$’’ token, ensuring stable and precise numeric generation.}

\begin{figure}[!h]
    \centering
    \includegraphics[width=0.6\linewidth]{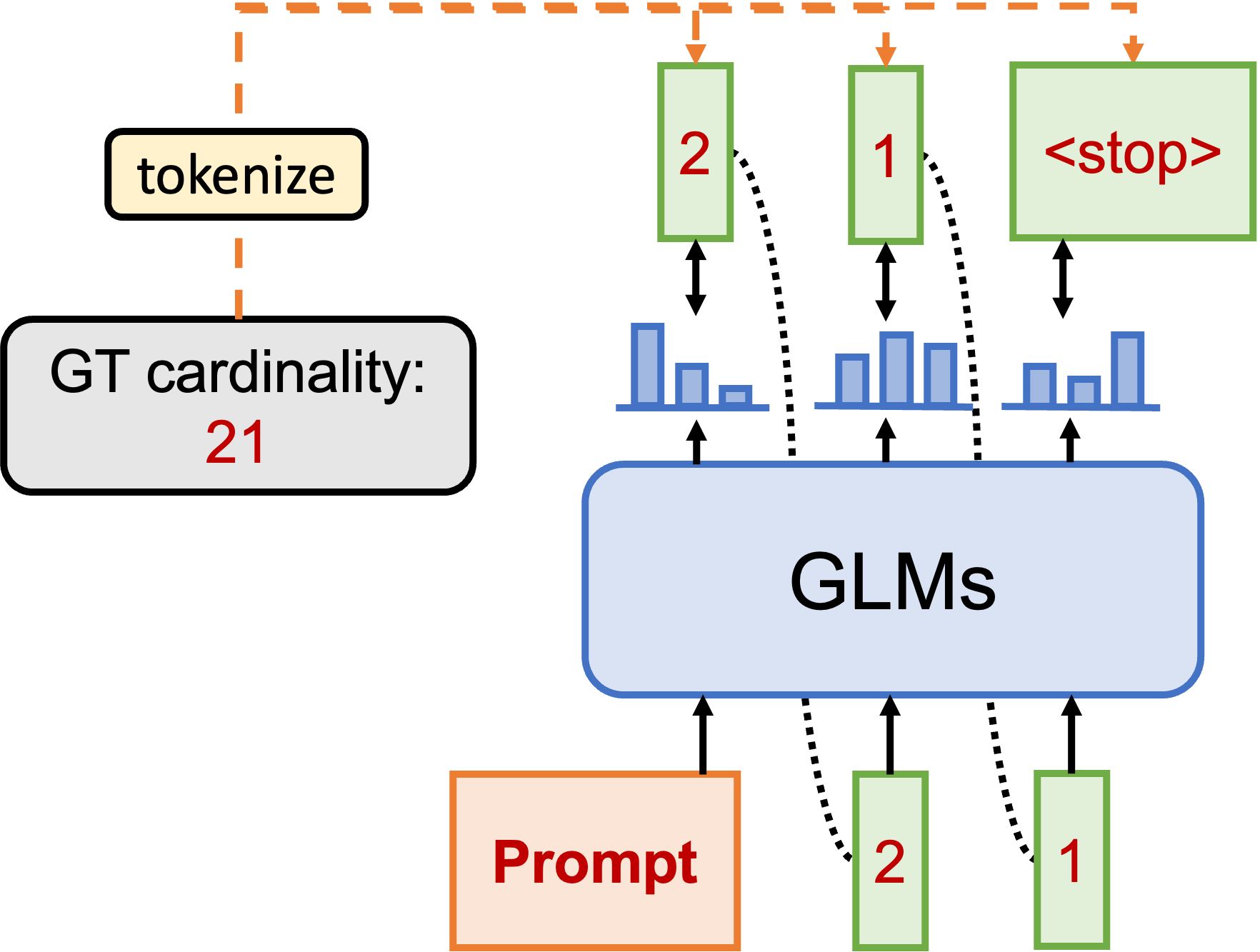}
    \caption{Fine-tuning \llm\ for \ce}
    \label{fig:train}
\end{figure}

\subsection{Fine-tuning \llm\ for \ce}\label{sec: fine_tune_llm}

Note that as mentioned in Section \ref{sec: related_work}, there is an emerging trend recently that \llm\ are increasingly adapted for database tasks such as knob tuning, database performance diagnosis and query rewrite. Most of these prior works keep \llm\ frozen and simply leverage the few-shot or zero-shot reasoning power of \llm. However, our initial exploration indicates that merely prompting frozen \llm\ may \reviewerfour{produce sub-optimal} \ce\ performance, even generating non-numeric token sequences most times. Hence, it becomes essential to fine-tune \llm\ so that they can serve for \ce\ task. To accomplish this, for each SQL query $Q$ in a training set $D$, we first tokenized its ground-truth cardinality into a numeric token sequence, denoted by $\mathbf{x} = [x_1,x_2,...,x_n]$ and aim to maximize the log-likelihood of generating all such sequences in $D$ with \llm, leading to the following objective function:

\begin{small}
\begin{align}\label{eq: objective}
\begin{split}
    & \sum\nolimits_{(Q,\mathbf{x}) \in D} log(P_{\theta}(\mathbf{x}| \text{prompt}(Q)))\\
    & = \sum\nolimits_{(Q,\mathbf{x}) \in D} \sum\nolimits_{i=1}^{n}log(P_{\theta}(x_i| \text{prompt}(Q), x_1,x_2,\dots,x_{i-1}))
\end{split}
\end{align}    
\end{small}
To compute the probability of a token sequence \\$x_1, x_2,\dots,x_n$ induced by the ground-truth cardinality of the query $Q$, the above formula first transforms this quantity into the product of the conditional probabilities of individual tokens $x_i (i=1,2,\dots,n)$, which employs the probability chain rule \cite{schum2001evidential}. Each conditional probability for token $x_i$ is then evaluated with an LLM parameterized by $\theta$, taking the prompt for $Q$ and the sub-sequence $x_1, x_2,\dots,x_{i-1}$ before $x_i$ as input.

This process is explained with one example query having ground-truth cardinality ``21'' in Figure \ref{fig:train}. As this figure suggests, the cardinality number ``21'' is firstly tokenized into multiple numeric digits and one additional terminating token, i.e., $x_1=2$, $x_2=1$ and $x_3=$$<$stop$>$. The generation probability of each token $x_i (i=1,2,3)$ is collected from the probability distribution of all tokens (including non-numeric ones) produced by \llm. This probability distribution is calculated by inputting the prompt proposed in Section \ref{sec: glm_prompt} and the tokens prior to $x_i$ to \llm.

\textit{Discussion}
It is worth noting that the above fine-tuning process differs from all the traditional ML-based cardinality estimators. Generally speaking, those traditional methods, in particular, those query-driven methods, first leverage neural networks such as transformer \cite{zeng2024price, li2023alece} or convolutional layers \cite{kipflearned} to encode queries and essential statistics collected from the underlying databases. These encodings are then fed into a multilayer perceptron (MLP) layer to output estimated cardinality numbers. In contrast, the above fine-tuning strategy aims to output a sequence of numeric tokens to compose the estimated cardinalities. In Section \ref{sec: ablation}, we attempt to follow the conventional methodology to employ one LLM as an encoder and one subsequent MLP layer to estimate cardinalities. The MLP layer and the LLM are jointly fine-tuned. However, this strategy exhibits a severe overfitting issue as indicated by our empirical results in Section \ref{sec: ablation}. This demonstrates the necessity of designing a novel paradigm for LLM-based cardinality estimators.

\textit{Pre-fine-tuning \llm\ across multiple pre-training datasets}
Recall that a desired property for \llm, particularly those fine-tuned with the above strategy, is its generalizability to unseen database instances. Considering the highly diverse database distributions and query patterns in the wild, we thus follow the setting of \cite{zeng2024price} to fine-tune \llm\ on 26 various pre-training datasets with the following objective function, which sum up Equation \eqref{eq: objective} over all datasets:

\begin{small}
\begin{align*}
    \sum_{D}\sum_{(Q,\mathbf{x}) \in D} \sum_{i=1}^{n}log(P_{\theta}(``x_i"| \text{prompt}(Q), ``x_1",``x_2",\dots,``x_{i-1}"))
\end{align*}    
\end{small}

Note that although the above formula is for fine-tuning \llm, it mimics the pre-training process of \cite{zeng2024price} and the resulting models can be further fine-tuned on the training split of target datasets. To distinguish these two fine-tuning processes,
the one leveraging the above formula is referred to as {\em pre-fine-tuning} throughout this paper, which adheres to the terminology of \cite{kangget}. Our experiments show that the pre-fine-tuning phase is optional, meaning direct fine-tuning on target datasets is possible. We thus further evaluate its effect throughout the experiments.

\subsection{Self-correction during inference}\label{sec: glm_inference}
As discussed in machine learning literature \cite{gekhman2024does,huang2023survey}, fine-tuned \llm\ tend to produce hallucination, e.g., producing extreme cardinality numbers or even non-numeric tokens in \ce\ task. This can occur especially on tasks like \ce\ which are not incorporated in the internal knowledge of \llm\ \cite{schulman2023reinforcement}.

To rectify such estimation errors of \llm\ in the context of out-of-distribution queries or data, we propose to perform self-correction \cite{madaan2024self} during the inference phase when necessary. As discussed in \cite{pan2023automatically, shinn2024reflexion, madaan2024self, manakul2023selfcheckgpt}, \llm\ are capable of producing higher-quality output after receiving external feedback on their initial output. Specifically, we monitor the difference between the estimated cardinality of \llm\ and that of the cardinality estimators selected for composing prompts in Section \ref{sec: glm_prompt}. Whenever this difference exceeds a pre-specified threshold, this feedback signal and estimated cardinality are then appended to the original prompts of \llm\ for self-correction, which follow the principle of \cite{madaan2024self}. This process can be repeated multiple times until the estimated cardinalities are good enough. Our experiments in Section \ref{sec: ablation} justify the efficacy of this design. 

\reviewerfour{It is noteworthy that the threshold used for determining whether to perform self-correction is crucial. To guarantee selecting an appropriate threshold, we perform a grid search. Specifically, we tune this threshold individually for each baseline method whose estimated cardinality is provided for the LLM prompt, as their estimation error distributions vary across models and databases. This ensures that the self-correction mechanism adapts appropriately to the characteristics of each baseline rather than relying on a single global threshold.}

%% file: exp_setup.tex
\section{Overall experimental setups}\label{sec: exp_setups}
As mentioned in Section \ref{sec: intro}, we aim to perform a comprehensive evaluation of the capability of \llm\ for the \ce\ task in this paper. This section describes the general experimental setups, covering hardware and software configurations, LLM fine-tuning and inference settings, dataset description, evaluation metrics, and varied \llm\ that we evaluated. 

\textit{Hardware and Software} 
We perform all of our experiments on a GPU machine with 8 A-800 GPUs, each of which has 80 GB of memory. In addition, this machine contains an Intel(R) Xeon(R) CPU and 512 GB CPU memory. To reduce the effort of performing fine-tuning and inference with open-sourced \llm\, we employ LLaMA-Factory \cite{zheng2024llamafactory}, a generic language model fine-tuning framework. This framework supports a series of state-of-the-art open-sourced \llm\ including \\ Llama\cite{dubey2024llama}, LLaVA\cite{liu2023llava}, and DeepSeek \cite{deepseekv2}. In addition, Postgres \cite{postgresql2020} is the underlying database engine, which serves to collect essential statistics for the \ce\ task and support the end-to-end evaluation. 

\textit{Large Language Models (\llm)} The default LLM that is employed for the \ce\ task is Llama-3 8B \cite{dubey2024llama} (8B means that it has 8 billion parameters) since it is a broadly used and state-of-the-art open-source LLM. But we also perform ablation studies in Section \ref{sec: ablation} to evaluate the power of various \llm\ on the \ce\ task, including
DeepSeek V2 \cite{deepseekv2}, another representative open-source LLM with 16 billion parameters. 
and GPT-4o \cite{openai2023gpt4o}, one state-of-the-art closed-source language model. Note GPT-4o has 200 billion parameters, substantially exceeding DeepSeek and Llama 3. Also, different from these two models, fine-tuning GPT-4o is conducted by submitting requests through the API provided by openAI\footnote{see the instruction at \url{https://platform.openai.com/docs/guides/fine-tuning/faq}}.

\textit{Pre-training datasets for pre-fine-tuning} We reuse the datasets that have been used for pre-training in \cite{zeng2024price} for performing pre-fine-tuning on \llm. These datasets are composed of 24 datasets assembled from \cite{motl2015ctu} and two benchmark datasets, TPC-H \cite{conf/ndt/Moussa12} and SSB \cite{o2009star}. The authors of \cite{zeng2024price} performed extensive cleaning and pre-processing on these datasets. They additionally generated synthetic workloads as described in \cite{zeng2024price} for the pre-fine-tuning phase, which is further discussed in Section \ref{sec: benchmark}. The resulting datasets exhibit very diverse data distributions, schema and sizes. These datasets span various domains like government, sports, retail, and medicine. Thus, these 26 datasets are regarded as a perfect surrogate for real-world databases in pre-fine-tuning \llm.


\textit{Inference and Fine-tuning settings} 
As mentioned above, we fine-tune GPT-4o and open-source \llm\ by leveraging API invocation and LLaMA-Factory \cite{zheng2024llamafactory} respectively. To accelerate fine-tuning open-source \llm, we leverage the Low-Rank Adaptation (LoRA) technique \cite{hulora} to only fine-tune a small portion of trainable parameters, which turns out an efficient way of fine-tuning \llm\ without sacrificing performance in general. In addition, we discover that it is sufficient to just fine-tune \llm\ for a few epochs so that \llm\ only output numeric tokens and produce small enough validation errors.
The batch size for fine-tuning is configured as 16, which can use up all 80GB of memory in one A-800 GPU. 

As mentioned in Section \ref{sec: fine_tune_llm}, we can optionally perform pre-fine-tuning on diverse pre-training datasets, followed by an additional optional fine-tuning phase on target datasets. In light of this, for each \llm, we evaluate three versions of it, i.e., LLM + FT, LLM + PFT, and LLM + PFT + FT respectively, in which LLM is an arbitrary language model, PFT represents performing pre-fine-tuning while FT represents fine-tuning. Recall that to facilitate accurate \ce, \llm\ needs to incorporate the cardinalities estimated by PRICE into its prompts. But to ensure fair comparison throughout the paper, whenever \llm\ are not fine-tuned on target datasets, i.e., LLM + PFT is used, we incorporate the cardinalities estimated by the pre-trained version of PRICE (denoted by PRICE w/o FT) into the prompts of \llm. Otherwise, the estimated cardinalities by the fine-tuned version of PRICE are used.  
As we will see in Section \ref{sec: data_efficiency}, the number of samples in pre-training datasets plays a critical role in the quality of the pre-trained \llm, To employ the pre-trained \llm\ for experiments except those in Section \ref{sec: data_efficiency}, we vary the size of each pre-training dataset to identify the smallest size such that the corresponding pre-trained \llm\ can almost match performance as the pre-trained PRICE model \cite{zeng2024price}, the state-of-the-art pre-trained \ce\ model. Eventually, this size is identified as 8000 per pre-training dataset and the details of determining this number are provided in Section \ref{sec: data_efficiency}. 


\textit{Evaluation metrics}
We leverage the following metrics to compare the effectiveness between methods on the \ce\ task, including:
\begin{enumerate}[noitemsep, topsep=0pt, left=0pt]
    \item \textbf{Q-error}\cite{moerkotte2009preventing} which computes the relative difference between the estimated cardinality $\hat{c}$ and the ground truth $c$, i.e., $\text{Q-error} = \max(\hat{c}/c, c/\hat{c})$. 
    By following prior works such as \cite{zeng2024price,li2023alece}, Q-errors can be evaluated at varied quantiles, which range from 50\% to 99\% throughout experiments. Intuitively speaking, larger-quantile Q-errors reflect the \ce\ performance on harder queries.
    \item \textbf{End-to-End time} which is the actual query execution time using the estimated cardinalities. We employ the implementation provided by \cite{zeng2024price} for evaluating this metric. 
\end{enumerate}



%% file: exp_benchmark.tex
\section{Performance on simple SPJ queries}\label{sec: benchmark}
We first follow \cite{zeng2024price} to evaluate the \ce\ performance of \llm\ on simple SPJ queries with four widely used benchmark datasets. It is then compared against conventional ML-based cardinality estimators. Since the training set and test set are sampled from the same distribution in those datasets, these experiments can suggest the {\em in-distribution generalizability} of \llm\ in the \ce\ task.

\subsection{Experimental setups}
\textit{Datasets}
All the experiments in the paper are performed on the following well-established benchmark datasets or their variants:
\begin{itemize}[noitemsep, topsep=0pt, left=0pt]
\item IMDB \cite{leis2015good} which is on movies and actors. Queries from the Job-light workload \cite{job_light_workload} on this dataset are used;
\item STATS \cite{stats_dataset, han2021cardinality} which consists of user-contributed content on the Stats Stack Exchange network. We perform the \ce\ on the queries from the STATS-CEB workload \cite{han2021cardinality} which primarily contains manually crafted multi-table join queries. 
\item ErgastF1 \cite{ergastF1Dataset} which is a collection of Formula 1 (F1) racing data maintained by the Ergast Developer API.
\item \vg\ \cite{VisualGenomeDataset} is a structured gene dataset. 
\end{itemize} 
Note that since no queries are provided in the ErgastF1 and \vg\ datasets, we thus follow \cite{zeng2024price} to generate synthetic queries. This query generation process starts by sampling a sub-graph from the join schema graph in those two datasets. Each sampled sub-graph represents a join query between tables appearing in this sub-graph, which is also subsequently associated with randomly produced predicates. Specifically, we randomly select a numeric or categorical column $A$ and constants from the domain of the column $A$ to compose a filter predicate, which is repeated $y$ times. Note that if $A$ is a numeric column, two constants, $l$ and $u$ are sampled to compose a filter $A \leq u$ and $A \geq l$. On the other hand, if $A$ is a categorical column, we only produce one constant $a$ to construct a filter $A=a$. In addition, we follow \cite{zeng2024price} to partition the query sets in the above four datasets into training and testing splits. In this section, the same set of training queries is used for fine-tuning \llm\ and training baseline query-driven methods for fair comparison.

\textit{Baseline methods} Recall that conventional ML-based cardinality estimators belong to three categories, i.e., data-driven, query-driven, and hybrid methods. Hence, we compare \llm\ against representative methods belonging to these three classes, including state-of-the-art data-driven methods, DeepDB \cite{hilprecht13deepdb}, NeuroCard \cite{yang2020neurocard}, FactorJoin \cite{wu2023factorjoin} and ASM \cite{kim2024asm}, one broadly used query-driven method, MSCN \cite{kipflearned}, two state-of-the-art hybrid method, ALECE \cite{li2023alece} and PRICE \cite{zeng2024price}. Note that our initial trials suggest that not all data-driven methods are compatible with all the above four datasets. For instance, 
DeepDB cannot be run on \ergast\ and \vg\ dataset. Similarly, NeuroCard is incompatible with \ergast, and the execution of FactorJoin fails on \ergast\ and \vg\ dataset. Hence, we skip the evaluation of these methods on those datasets. Also, PRICE \cite{zeng2024price} constructs a pre-trained model, which can be employed directly for estimating cardinalities on arbitrary datasets or be further fine-tuned on target datasets. In this experiment, both versions are used for comparison, which are denoted by PRICE (w/o FT) and PRICE respectively. In addition to the above ML-based methods, we also include the default cardinality estimator in PostgreSQL as one baseline, which is denoted by PG. \reviewertwo{Moreover, we include PRICE+PG, a selective estimation variant that utilizes PG for low-cost queries and PRICE for high-cost queries, which is further explained in Section~\ref{sec: e2e_ft}.}

\subsection{Experimental results}\label{sec: main_results}
We present the results in Table \ref{Table: over_accuracy}. Note that due to the lack of essential statistical information and memory leakage issues, we were unable to obtain results for NeuroCard and DeepDB on datasets other than IMDB and STATS, which are thus not reported in this table. This challenge has been documented in previous works such as \cite{li2023alece} and \cite{zeng2024price}.

According to Table \ref{Table: over_accuracy}, one observation is that fine-tuning (both the raw version and the pre-fine-tuned version of) \llm\ achieves better estimation accuracy than all the baseline methods across nearly all datasets and percentile Q-errors. Notably, the reduction of Q-errors w.r.t. the best baseline method is up to 74.1\% (see 99-percentile Q-error on \stats\ dataset). On the other hand, other methods, in particular, those state-of-the-art, exhibit variability across datasets and even across varied percentile Q-error within the same dataset. For instance, the low-quantile Q-error of ALECE is generally smaller than that of PRICE while PRICE\footnote{Note that the performance numbers of PRICE that we report deviate from those in \cite{zeng2024price}. This is because in \cite{zeng2024price}, the Q-error is calculated on the decomposed sub-queries of each query rather than the entire query from the test set for PRICE. The latter option is employed in this paper across all methods for fair comparison} can beat ALECE on high-percentile Q-errors. Similarly, NeuroCard performs consistently better than ALECE on IMDB while incurring much higher Q-errors than ALECE on STATS. 

It is worth noting that the performance of Llama + PFT + FT surpasses all other methods, including that of Llama + FT. This thus demonstrates the benefit of pre-fine-tuning \llm\ to absorb the prior knowledge on cardinality estimation from diverse pre-trained datasets. 
On the other hand, without further fine-tuning, pre-fine-tuned LLM, i.e., Llama + PFT,
outperforms the pre-trained PRICE model in almost all cases while they still produce higher estimation errors than many other methods. However, it is worth noting that Llama + PFT estimates more accurately than many baseline methods such as ALECE, PRICE, MSCN, and PG in IMDB. Considering that no fine-tuning on these four datasets is performed for Llama + PFT , this thus indicates the remarkable generalizability of pre-fine-tuned \llm\ to unseen datasets. Note that only 8000 queries pre-training dataset are employed for pre-fine-tuning, which is only around 16\% of the dataset size used for pre-training PRICE \cite{zeng2024price}. As we will discuss in Section \ref{sec: data_efficiency}, the estimation errors of the pre-fine-tuned \llm\ on unseen datasets stably decrease with the growing sizes of pre-training datasets. Therefore, with larger pre-training datasets, better \ce\ performance for Llama + PFT would be expected. 

We also reported the training time for each method, which is highlighted in green in Table \ref{Table: over_accuracy}. For Llama + PFT, its training process is indeed the pre-fine-tuning process and thus the pre-fine-tuning time is reported. In contrast, for Llama + PFT + FT and Llama + FT, we report the fine-tuning time. Although it takes up to around 12 hours to pre-fine-tune Llama, it is a one-time and offline process and the resulting model can be applied to arbitrary database instances without repeating the pre-fine-tune process. Building on top of it, additional fine-tunings can be conducted for further performance enhancement on any target database, which only takes around 50 to 60 mins. As suggested by Table \ref{Table: over_accuracy}, this fine-tuning overhead is comparable to most baseline methods, and even smaller than that of DeepDB, FactorJoin, and ALECE. Such fast fine-tuning time on Llama can be attributed to multi-GPU setups and the application of the LoRA (Low-Rank Adaptation) method. Therefore, applying LLMs such as Llama to \ce\ brings more accurate estimation results without incurring significant training overhead. It is worth noting that although other baseline methods, such as NeuroCard, ASM, and PRICE, generally incur shorter training time, their \ce\ performance is worse than Llama. 




\subsection{Main findings}
Our findings of this set of experiments are summarized below:
\begin{itemize}[noitemsep, topsep=0pt, left=0pt]
\item Fine-tuning is effective in empowering (both the raw version and the pre-fine-tuned version of) \llm\ to beat the state-of-the-art cardinality estimators. Larger performance gains brought by \llm\ typically occur on those long queries. 
\item In comparison to fine-tuning raw \llm, fine-tuning the pre-fine-tuned \llm\ results in more accurate estimations on cardinalities, thus demonstrating the necessity of the pre-fine-tuning phase in enhancing the \ce\ performance. 
\item Even without any fine-tuning overhead on target \\datasets, the pre-fine-tuned \llm\ still outperform many baseline methods in many cases, thus highlighting their generalizability power to unseen databases.
\item Both the pre-fine-tuning and fine-tuning incur reasonable overhead. Despite longer running time, the former is a one-time and offline phase and the resulting model can be applied to arbitrary database instances. In contrast, fine-tuning needs to be repeated on each database instance but only takes less than one hour, which is comparable to most baseline methods. 

\end{itemize}

\begin{table*}[!h]
\small
    \centering
    \resizebox{\textwidth}{!}{
\begin{tabular}{c|cccc|cccc|cccc|cccc}\toprule
\multirow{3}{*}{}&\multicolumn{4}{c|}{\imdb}&\multicolumn{4}{c|}{\stats}&\multicolumn{4}{c|}{\ergast}&\multicolumn{4}{c}{\vg}
\\
& \multicolumn{3}{c}{Q error} & \multirow{2}{*}{
\begin{tabular}{c}
Training time\\(mins)
\end{tabular}
} & \multicolumn{3}{c}{Q error} & \multirow{2}{*}{
\begin{tabular}{c}
Training time\\(mins)
\end{tabular}
} & \multicolumn{3}{c}{Q error} & \multirow{2}{*}{
\begin{tabular}{c}
Training time\\(mins)
\end{tabular}
} & \multicolumn{3}{c}{Q error} & \multirow{2}{*}{
\begin{tabular}{c}
Training time\\ (mins)
\end{tabular}
} \\ 
 & 50\% &90\% & 99\%& & 50\% &90\% & 99\%& & 50\% &90\%& 99\%& & 50\% &90\%& 99\% & \\\hline
\fbox{PG} & 1.95&19.04&{1043.54}&\cellcolor{green!25}-&1.87&20.71&{1765.71}&\cellcolor{green!25}-&1.60&11.30&114.24&\cellcolor{green!25}-& 1.17  & 15.12 & {11711.42}&\cellcolor{green!25}- \\ \midrule
MSCN&3.24&28.54&411.90&\cellcolor{green!25}75.19&6.19&364&{43608.31}&\cellcolor{green!25}2.03&10.20&353&{4658.09}&\cellcolor{green!25}5.08& 2.84  & 18.91 & 108.48&\cellcolor{green!25}38.86\\ \midrule
DeepDB &\textbf{1.36}&5.62&26.45&\cellcolor{green!25}73.73&1.84&73.52&1507.78&\cellcolor{green!25}142.12&-&-&-&\cellcolor{green!25}-&-&-&-&\cellcolor{green!25}-\\
NeuroCard &1.66&7.80&27.95&\cellcolor{green!25}14.55&2.12&48.40&{1035.66}&\cellcolor{green!25}36.43&-&-&-&\cellcolor{green!25}-&1.04&14.35&57.35&\cellcolor{green!25}12.91\\
FactorJoin &13.86&348.96&{5979.40}&\cellcolor{green!25}78.45&5.41&150.52&{8326.57}&\cellcolor{green!25}0.55&-&-&-&\cellcolor{green!25}-&-&-&-&\cellcolor{green!25}-\\
ASM &{5.99}& {79.21}& {760.30}&\cellcolor{green!25}1000.6& {8.62}&{454.99}&{60523.63}&\cellcolor{green!25}2.15&{46.41}&{459565}&{112039519}&\cellcolor{green!25}1.40&{1.58} &{101.97} &{2995.50}&\cellcolor{green!25}5.95 \\ \midrule
ALECE &1.75&11.49&124.68&\cellcolor{green!25}472.88&1.67&7.93&119&\cellcolor{green!25}93.09& 1.75  & 5.18& 53.81&\cellcolor{green!25}121.49& 1.13 & 1.44 & 2.01&\cellcolor{green!25}49.68\\
\fbox{PRICE (w/o FT)} &4.02&60.25&186&\cellcolor{green!25}313.56 &4.98&110&{9654.00}&\cellcolor{green!25}313.56& 1.42  & 11.06  & 129&\cellcolor{green!25}313.56& 3.63  & 5.32 & 62.66&\cellcolor{green!25}313.56\\
PRICE &2.26&13.09&98.91&\cellcolor{green!25}6.47&2.91&33.31&697&\cellcolor{green!25}10.20& 1.81  & 6.89 & 26.95&\cellcolor{green!25}15.21& 1.54 &  2.82 & 13.68&\cellcolor{green!25}11.48\\
{PRICE+PG} & {2.13}&{16.57}&{645.88}&\cellcolor{green!25}{6.47}&{2.32}&{25.43}&{1414.58}&\cellcolor{green!25}{10.20}& {1.73}  &{9.86} & {67.02}&\cellcolor{green!25}{15.21}& {1.27} &{7.38} &{1324.16} &\cellcolor{green!25}{11.48}\\ 
\midrule
Llama + FT &1.64&9.26&35.85&\cellcolor{green!25}47.91&1.35&7.76&33.04&\cellcolor{green!25}53.84& 1.37  & 3.17 & 47.36&\cellcolor{green!25}61.28& 1.01  & 1.19 & 1.59&\cellcolor{green!25}44.32\\
\fbox{Llama + PFT} &1.65&6.24&26.53&\cellcolor{green!25}704.31& 2.19 & 6.72 & 198.55&\cellcolor{green!25}704.31& 2.03  & 8.35 & 95.19&\cellcolor{green!25}704.31& 1.56  & 8.49 & 38.53&\cellcolor{green!25}704.31\\
Llama + PFT + FT &1.60&\textbf{5.42}&\textbf{21.06}&\cellcolor{green!25}47.91&\textbf{1.33}&\textbf{5.77}&\textbf{30.90}&\cellcolor{green!25}53.84& \textbf{1.19} & \textbf{2.28} & \textbf{16.03}&\cellcolor{green!25}61.28& \textbf{1.00} & \textbf{1.13} & \textbf{1.47}&\cellcolor{green!25}44.32\\
\bottomrule
\end{tabular}
}
\caption{Overall performance on different \ce\ benchmark datasets. The methods highlighted with boxes correspond to the ones that are not fine-tuned on these benchmark datasets. Hence, for those methods, we report their pre-training or pre-fine-tuning times in the training time column. 
}\label{Table: over_accuracy}
\end{table*}

%% file: exp_generalization.tex
\section{{Out-Of-Distribution Generalizability}}\label{sec: generalizability}
We study the \llm' out-of-distribution generalizability in this section, including the generalizability to new query patterns (i.e., workload shift, see Section \ref{sec: gen_workload}), data updates (Section \ref{sec: gen_data_dist})\reviewerone{, data distribution shift (Section \ref{sec: gen_data_distribution})} and unseen database instances (Section \ref{sec: gen_unseen_db}). 

\subsection{Generalizability to workload shift}\label{sec: gen_workload}
\subsubsection{Experimental setups}
We follow \cite{zeng2024price} to utilize \imdb\ \reviewertwo{and \stats}\ dataset for this experiment. But unlike Section \ref{sec: benchmark} which employs the Job-light workload, we design a synthetic workload to emulate the shift from one set of query templates to another disjoint set of queries with different templates. Specifically, by following the query generation process mentioned in Section \ref{sec: benchmark} for the ErgastF1 and VisualGenome dataset, we randomly produce queries with fewer than 3 joins as the training set for fine-tuning and another set of queries with more than 3 joins as the test set. Likewise, we vary the number of filters by producing queries with less than 4 filter predicates and queries with more than 4 filter predicates as the training set and test set, respectively. The experiments with these two different training and test sets are referred to as the {\em join query experiments} and {\it filter query experiments} hereafter. The training and test set in these two experiments contain around 40000 and 3000 queries, respectively.

{To further enhance the model’s robustness to workload shift, we additionally generate a new set of queries using the state-of-the-art LLM-based SQL query generation method \cite{lao2025sqlbarber}, which is then employed to augment the training set used for the pre-fine-tuning stage. These queries are structurally more complex than those in the original pre-training datasets, yet simpler than the queries used in the test workloads, meaning that each of these new queries contains no more than 3 joins and 4 filters. The pre-fine-tuning process using these new datasets is denoted as {$PFT^{comp}$} in this section. } {The reasons of augmenting the training set to include these new types of queries is that zero-shot generalization to unseen query types is impossible unless these query types are observed in the training corpus. Indeed, prior work shows that LLM generalization is fundamentally constrained by the coverage and diversity of the training corpus. Empirical studies indicate that models generalize to ``unseen'' tasks largely via interpolation over previously observed semantic and structural patterns, rather than true extrapolation to entirely novel query types \cite{wettig2024qurating, myntti2025register}. Moreover, analyses of zero-shot performance suggest that apparent zero-shot gains often arise from latent overlap between evaluation tasks and pretraining data \cite{zhang2024regurgitative,hoffmann2022training}. Consequently, robust performance on unseen query types generally requires that related patterns be present in the training corpus.}

The configurations of \llm\ remain consistent with Section \ref{sec: benchmark}. In this experiment, we compare \llm\ against MSCN and DeepDB, the best-performing query-driven and data-driven methods on \imdb. We also include PRICE rather than ALECE into the baseline methods since PRICE outperforms ALECE under workload shift as suggested by \cite{zeng2024price}. \reviewertwo{Since DeepDB shows relatively poor performance on the STATS dataset (as reported in Table \ref{Table: over_accuracy}), we omit its results for this dataset in this experiment.}

{To further evaluate the generalization ability of our method under a more rigorous OOD protocol, we additionally conduct experiments following a cross-dataset OOD setting as used in \cite{negi2023robust}. Specifically, we fine-tune all models on the CEB dataset and evaluate them on the JOB-light workload, ensuring that the query templates and data distributions between the training and testing sets are entirely disjoint. This setting better reflects real-world workload shift scenarios. We compare LLM-based methods with PG and PRICE. DeepDB and MSCN are excluded in this experiment since they are inapplicable to such cross-dataset OOD settings.
}

\subsubsection{Experimental results}\label{sec: gen_workload_shift_result}
The experimental results are included in Table \ref{Table: workload_shift_accuracy} {and Table \ref{Table: workload_shift_accuracy_stats}}. As Table \ref{Table: workload_shift_accuracy} indicates, \llm\ consistently produce smaller Q-error than all baseline methods at all percentiles. Notably, in the join query experiments, in comparison to the best baseline, PRICE, \llm\ bring more significant performance gains with increasing quantiles, which can achieve up to 75.81\% Q-error reduction (10672.53 VS 2581.73) on 99 percentile cases. This thus demonstrates that \llm\ are more adept at generalizing to unseen join query patterns, in particular, those hard ones, than the baseline methods. 
On the other hand, in the filter query experiments, the relative performance improvement of \llm\ over the baseline methods remains stable across all percentile Q-errors, which is around 30\%-40\%. Note that DeepDB, the best-performing data-driving method on \imdb\ dataset as indicated by Table \ref{Table: over_accuracy} in Section \ref{sec: benchmark}, exhibits over one order-of-magnitude higher Q-errors than \llm. This occurs despite its theoretical robustness to workload shift. 

{In Table \ref{Table: workload_shift_accuracy} and Table \ref{Table: workload_shift_accuracy_stats}}, the slightly worse 99th-percentile q-error of Llama+PFT+FT compared to Llama+FT mainly stems from the fact that the PFT stage tends to overfit simple query patterns, leading to reduced robustness on complex queries. However, under the {$PFT^{comp}$} setting, \llm\ are exposed to more complex query patterns during the pre-fine-tuning stage, which enables them to better adapt to dynamic workloads and further reduce the overall Q-error compared to the standard PFT configuration.

{As reported in Table \ref{Table: workload_shift_accuracy_stats}. In the join-heavy workload ($>$3 joins), Llama+{$PFT^{comp}$}+FT achieves a 99th-percentile Q-error of 2906.43, which is significantly lower than the 11943.00 of PRICE and 20135.00 of MSCN. Likewise, for filter-heavy queries, our method maintains its lead with a 50th-percentile Q-error of 3.23, outperforming both PRICE and PG. These results across both IMDB and STATS datasets further solidify the conclusion that LLMs possess stronger robustness to workload shifts than existing methods.}

{Furthermore, for the more rigorous OOD protocol, the results in Table \ref{Table: ood_job} show that Llama+PFT+FT consistently achieves the lowest Q-error among all compared methods, highlighting its strong generalization ability under workload shifts.}

\begin{table}[h]
\centering
    \resizebox{0.5\textwidth}{!}
    {
\begin{tabular}{c|c|cccc}\toprule
\multirow{2}{*}{Workload}& \multirow{2}{*}{\ce\ method}&\multicolumn{4}{c}{Q-error} 
\\
& & 50\% & 90\% & 95\% & 99\% \\\hline
\multirow{9}{*}{
\begin{tabular}{c}
Train \\ ($<$ 3 joins), \\ Test \\ ($>$ 3 joins)
\end{tabular}
} & PG & 30.48 & 1344.01 & 4939.27 & 52594 \\
& MSCN & 21.77 & 387.61 & 2217.41 & 17239\\
& DeepDB & 98.67 & 2376.85 & 4270.22 & 20236\\
& PRICE & 11.90  & 392.93 & 857.23 & 10672 \\ \hhline{~-----}
& Llama + FT & 9.18  & 131.92 & 323.67 & 2581.73\\
& Llama + PFT & 14.79  & 521.47 & 1255.56 & 14247\\
& {Llama + {$PFT^{comp}$}} & {12.64} & {423.59} & {683.22} & {9371.08} \\
& Llama + PFT + FT & 9.10 & 93.34 & 315.64 & 7164.52 \\
& {Llama + {$PFT^{comp}$} + FT} & {\textbf{8.98}} & {\textbf{87.46}} & {\textbf{303.17}} & {\textbf{2438.91}}\\
\hline
\multirow{9}{*}{
\begin{tabular}{c}
Train \\($<$ 4 filters), \\ Test \\ ($>$ 4 filters)
\end{tabular}
} & PG & 9.71  & 510.72 & 2094.89 & 15307\\
& MSCN & 7.94 & 88.64 & 551.92 & 7843.92\\
& DeepDB & 42.37 & 7098.95 & 51424 & 82385\\
& PRICE & 5.02  & 97.58 & 323.89 & 4276.47\\\hhline{~-----}
& Llama + FT & 3.49  & 53.81 & 211.37 & 3233.39\\
& Llama + PFT & 5.41  & 118.26 & 377.49 & 3760.76\\
& {Llama + {$PFT^{comp}$}} & {4.36}  & {71.92} & {265.22} & {3355.81}\\
& Llama + PFT + FT & 2.26 & 28.69 & 139.58 & 2092.42 \\
& {Llama + {$PFT^{comp}$} + FT} & {\textbf{2.03}} & {\textbf{25.54}} & {\textbf{123.08}} & {\textbf{1952.16}}\\
\bottomrule
\end{tabular}
}
\caption{\reviewertwo{Performance on \imdb\ dataset under workload shift} 
}\label{Table: workload_shift_accuracy}
\end{table}

\begin{table}[h]
\small
    \centering
    \resizebox{0.5\textwidth}{!}{
\begin{tabular}{c|c|cccc}\toprule
\multirow{2}{*}{Workload}& \multirow{2}{*}{\ce\ method}&\multicolumn{4}{c}{Q-error} 
\\
& & 50\% & 90\% & 95\% & 99\% \\\hline

\multirow{10}{*}{
\begin{tabular}{c}
Train \\ ($<$ 3 joins), \\ Test \\ ($>$ 3 joins)
\end{tabular}
} & PG & 42.05 & 1507.11 & 4546.31 & 50012 \\
& MSCN & 29.64 & 438.11 & 2678.09 & 20135\\
& {DeepDB} &{27.61}  & {503.18} &{3144.70}  &{19423} \\
& {Neurocard} & {27.93} &{401.28}  &{2024.77}  &{14808} \\
& PRICE &  13.58 & 464.15 & 899.23 & 11943 \\ \hhline{~-----}
& Llama + FT &  11.67  & 145.03 & 409.58 & 2937.99\\
& {Llama + PFT} &{26.35}  & {701.23} &{1644.89}  &{17306} \\
& Llama + {$PFT^{comp}$} & 18.34  & 572.01 & 754.15 & 9570.13 \\
& {Llama + PFT + FT} &{11.41}  & {119.07} &{403.29}  &{8386.52} \\
& Llama + {$PFT^{comp}$} + FT & \textbf{10.95} & \textbf{105.67} & \textbf{367.92} &\textbf{2906.43}  \\
\hline
\multirow{6}{*}{
\begin{tabular}{c}
Train \\($<$ 4 filters), \\ Test \\ ($>$ 4 filters)
\end{tabular}
} & PG & 10.23  & 398.79 & 1955.03 &12087 \\
& MSCN & 8.58 & 73.74& 569.05&7323.55\\
& {DeepDB} &{7.67}  & {132.74} &{540.73}  &{6956.09} \\
& {Neurocard} & {8.19} &{63.21}  &{488.53}  &{6084.44} \\
& PRICE &6.23 &88.01 &309.46 &4001.84\\\hhline{~-----}
& Llama + FT & 3.97 &46.65 &192.21 &2958.71\\
& {Llama + PFT} &{6.11}  & {92.34} &{298.75}  &{3924.08} \\
& Llama + {$PFT^{comp}$} &5.41 &89.66 &252.08 &3617.32\\
& {Llama + PFT + FT} &{3.57}  & {40.26} &{120.99}  &{2279.63} \\
& Llama + {$PFT^{comp}$} + FT &\textbf{3.23} &\textbf{31.51} &\textbf{114.90} & \textbf{2054.37}\\
\bottomrule
\end{tabular}
}
\caption{Performance on \stats\ dataset under workload shift 
}\label{Table: workload_shift_accuracy_stats}
\end{table}

\begin{table}[!h]
\small
    \centering
\begin{tabular}{c|cccc}\toprule
\multirow{2}{*}{\ce\ method}&\multicolumn{4}{c}{Q-error}\\
& 50\%  & 90\% & 95\% & 99\%  \\\hline
 PG & {1.95} & {19.04} & {49.44} & {1043.54} \\
PRICE & 16.07 & 45.66 & 54.34 & 75.89 \\
Llama + FT & {1.93} & {11.57} & {39.10} & {47.38} \\
Llama + PFT + FT & {\textbf{1.71}} & {\textbf{7.93}} & {\textbf{16.78}} &{\textbf{28.54}} \\
\hline
\end{tabular}
\caption{\reviewertwo{Performance on JOB-Light workload under workload shift}}\label{Table: ood_job}
\end{table}

\subsection{Generalizability to data updates}\label{sec: gen_data_dist}
\subsubsection{Experimental setups}\label{Sec:data_updates_setup}
We follow \cite{li2023alece} to evaluate the generalizability of \llm\ under write-intensive workloads on \stats\ dataset. 
To emulate the dynamic workloads in practice, we initialize the experiments with 2/3 of records in \stats\ dataset and subsequently perform insertion, updates, deletes and queries on this dataset. One insertion statement aims to add one of the remaining 1/3 records into \stats\ while one update or delete statement only impacts one random record. These statements are randomly shuffled with the queries from the STATS-CEB workloads. We then follow \cite{li2023alece} to craft two types of write-intensive workloads \reviewertwo{as fine-tuning datasets}, including the insert-heavy and update-heavy ones. The only difference between these two workloads is the ratios between the numbers of the insert, delete, and update statements, which are configured as 2:1:1 and 1:1:2 respectively. 

We include PG, MSCN \cite{kipf2019estimating}, Neurocard \cite{yang2020neurocard}, and ALECE \cite{li2023alece} as the baseline, in which the latter three are the best-performing query-driven, data-driven and hybrid method on \stats\ according to Table \ref{Table: over_accuracy}. 

\reviewerone{Note that since PRICE does not support dynamic workloads, we thus incorporate the estimates from ALECE, the most accurate \ce\ methods for such workloads so far, rather than PRICE, into the prompt for \llm. }

\subsubsection{Experimental results}\label{sec:gen_data_updates_results}
The experimental results are summarized in Table \ref{Table: write-intensive}. As this table suggests, Llama+FT is capable of reducing the estimation errors with respect to the baseline methods (see the bolded numbers in the lines for Llama+FT). The performance gains of \llm\ are more substantial under insert-heavy workloads, which reduces Q-error by up to 20\% (see 95-percentile Q-error). In contrast, \llm\ only achieve marginal performance gains in comparison to baseline methods under update-heavy workloads. Note that similar to the performance trends discovered in \cite{li2023alece}, except ALECE whose performance is on par with \llm, other methods produce orders-of-magnitude worse estimation errors than \llm\ and ALECE, thus demonstrating their poor generalizability to data updates on databases. 

Additionally, by comparing different versions of \llm, we observe that the pre-fine-tuned Llama, i.e., Llama+PFT struggles with significantly higher estimation errors than Llama+FT. Even after performing fine-tuning on pre-fine-tuned Llama, the resulting \ce\ performance is still worse than Llama+FT. This exhibits a totally different performance pattern from that in Section \ref{sec: benchmark} where the \ce\ performance can benefit from the pre-fine-tuning process. The ineffectiveness of pre-fine-tuned \llm\ in this experiment can be attributed to the lack of diverse workloads during the pre-fine-tuning phase which only focuses on read-only workloads. {In contrast, the FT stage is performed on datasets containing dynamic workloads that mix INSERT, UPDATE, and DELETE queries, allowing the \llm\ to better capture the characteristics of write-intensive scenarios. As a result, Llama+FT outperforms Llama+PFT. Moreover, since the PFT stage introduces a distribution shift due to its read-only training data, subsequent fine-tuning on write-intensive workloads cannot fully compensate for this mismatch, leading to the inferior performance of Llama+PFT+FT compared with Llama+FT.}

{To perform experiments in the context of data updates, we follow the setting in Section \ref{Sec:data_updates_setup}
to construct two types of write-intensive workloads with different ratios on the pre-training datasets, which are then used for pre-fine-tuning \llm. This pre-fine-tuning process involving dynamic workloads is denoted as {$PFT^{dyn}$} in Table \ref{Table: write-intensive}. By performing {$PFT^{dyn}$} on such dynamic workloads, \llm\ become more familiar with write-intensive workloads. After performing the same FT process, we find that Llama+{$PFT^{dyn}$}+FT slightly outperforms Llama+FT. This demonstrates that incorporating pre-fine-tuning on dynamic, write-intensive workloads can further enhance the LLM’s performance under distribution shifts and also highlights its ability to generalize under data updates.}

{
Furthermore, to verify the robustness of \llm\ without relying on strong external estimators like ALECE (which may not support all query types), we introduce a comprehensive PFT process denoted as $PFT^+$. This process incorporates both dynamic operations and complex query patterns (e.g., LIKE, DISTINCT) into the training corpus while strictly utilizing only coarse-grained PostgreSQL estimates in the prompt. As reported in Table \ref{Table: write-intensive}, Llama+$PFT^+$+FT achieves performance comparable to Llama+$PFT^{dyn}$+FT, and even outperforms it at the 99th percentile. This confirms that \llm\ can effectively interpret data shifts and maintain high accuracy using only coarse-grained statistics, eliminating the strong dependency on specific external models.
}

\begin{table}[!h]
\small
    \centering
    \resizebox{0.5\textwidth}{!}{
\begin{tabular}{c|c|ccccc}\toprule
\multirow{2}{*}{Workloads}& \multirow{2}{*}{\ce\ method}&\multicolumn{4}{c}{Q-error} 
\\
& & 50\% & 90\% & 95\% & 99\% \\\hline
\multirow{11}{*}{{Insert-Heavy}} & PG & 194.67 & {1326.54} & {2763.41} & {6798.03} \\
& Neurocard & 17.37 & {2032.50} & {2545.07} & {4352.19}\\
& MSCN & 20.09 & {1103.45} &{3697} & {4896.58} \\
& ALECE & 1.77 & 9.12 & 21.63 & 123.14\\ \hhline{~-----}
& Llama + FT & 1.54 & 8.95 &  17.32 & 112.90\\
& Llama + PFT & 1.97 & 15.01 &  40.50 & {2706.31}\\
& {Llama + {$PFT^{dyn}$}} & {1.68} & {13.72} &  {35.28} & {1636.80}\\
&{Llama + $PFT^+$} &{1.72} & {14.18}&{33.79} & {1188.60}\\
& Llama + PFT + FT & 1.91 & 10.50 &  23.25 & 145.49\\
& {Llama + {$PFT^{dyn}$} + FT} & {\textbf{1.51}} & {\textbf{8.63}} &  {\textbf{15.78}} & 101.91\\
&{Llama + $PFT^+$ + FT} &{1.60} & {9.41}&{15.93} & \textbf{{98.82}}\\
\hline
\multirow{11}{*}{{Update-Heavy}} & PG & 208.77 & {1564.72} & {3801.02} & {11201.46} \\
& Neurocard & 18.49 & {1005.70}& {2306.79} & {4783.44} \\
& MSCN & 18.89 & {1010.38} & {4002.10} & {10103.55}\\
& ALECE & 1.79 & 7.67 & 14.50 & 73.24\\ \hhline{~-----}
& Llama + FT & 1.76 & 7.24 & 14.01 & 63.88\\
& Llama + PFT & 3.55 & 14.48 &  24.17 & 140.24\\
& {Llama + {$PFT^{dyn}$}} & {2.79} & {10.96} & {16.63} & {79.03}\\
&{Llama + $PFT^+$} &{3.12} & {11.07}&{17.82} & {80.31}\\
& Llama + PFT + FT & 1.95 & 9.66 &  17.63 & 73.29\\
& {Llama + {$PFT^{dyn}$} + FT} & {\textbf{1.47}} & 5.26 &  {\textbf{10.03}} & 49.59\\
&{Llama + $PFT^+$ + FT} &{1.58} & \textbf{{5.01}}&{12.19} & \textbf{{45.67}}\\
\bottomrule
\end{tabular}
}
\caption{{Performance on \stats\ dataset under write-intensive workloads} 
}\label{Table: write-intensive}
\end{table}

\reviewerone{\subsection{Generalizability to data distribution shift}\label{sec: gen_data_distribution}}

\reviewerone{In this section, we investigate whether the models can generalize well when the underlying data distribution shifts while the database schema remain fixed. Specifically, we focus on evaluating Llama+PFT since it reflects the generalizability of LLMs from pre-trained datasets to unseen distributions. Among the baselines, PRICE is the only method that can be directly applied to new data distributions without fine-tuning. To this end, we evaluate PG, PRICE, and Llama+PFT on the TPC-H-Skew dataset\footnote{\url{https://github.com/YSU-Data-Lab/TPC-H-Skew}}, which introduces artificial data skewness compared with the uniform distribution in the standard TPC-H benchmark\footnote{\url{https://www.tpc.org/tpch/}}. 
As mentioned in Section \ref{sec: exp_setups}, the training datasets used in the pre-fine-tuning phase include the standard TPC-H benchmark. Hence, the experimental setting in this section simulates a real-world database instance with evolving data distributions over time.
} 

{As shown in Table \ref{Table: distribution_shift}, Llama+PFT achieves the lowest Q-errors at the 50th and 90th percentiles, demonstrating its robustness and adaptability to moderate data distribution shifts. However, at the 95th and 99th percentiles, its performance is comparable to that of PRICE. This can be attributed to the fact that the prompts for Llama+PFT incorporate the cardinality estimates of both PG and PRICE as auxiliary inputs. Consequently, when encountering queries that are highly sensitive to extreme data skewness, the model may partially inherit the estimation biases of these base estimators, limiting further improvements at the higher quantiles. Nevertheless, the overall results still indicate that Llama+PFT generalizes well to unseen data distributions and maintains competitive accuracy under significant distribution shifts.
}

\begin{table}[!h]
\small
    \centering
{ 
\begin{tabular}{c|cccc}
\toprule
\multirow{2}{*}{\ce\ method} & \multicolumn{4}{c}{Q-error} \\
& 50\% & 90\% & 95\%  & 99\%\\ \hline
PG & 1.78 & 299.75 & 354.23 & 415.38\\
PRICE & 1.65 & 30.33 & \textbf{89.54} & 167.40 \\
Llama + PFT & \textbf{1.60} & \textbf{9.588} & 93.71 & \textbf{163.52} \\ \hline
\end{tabular}
}
\caption{Performance on TPC-H-Skew datasets under data distribution shift}\label{Table: distribution_shift}
\end{table}

\subsection{Generalizability to unseen database instances}\label{sec: gen_unseen_db}
Indeed, the generalizability of \llm\ can be reflected with the results of \llm\ + PFT since \llm\ are pre-fine-tuned on 24 pre-training datasets and directly applied for computing cardinalities on target databases. In addition, among the baseline methods, only PRICE is capable of being applied to unseen databases without further fine-tuning. Hence, we only compare Llama+PFT against PRICE and PG for this experiment, and the results are included in the rows highlighted with boxes in Table \ref{Table: over_accuracy}. As these rows suggest, in comparison to PG and PRICE, Q-error gets reduced by up to around 67\% (see the 90\% Q-error on the IMDB dataset). In addition, the benefits brought by Llama+PFT are more significant on higher-quantile Q-errors than those on lower-quantile ones. This thus may indicate that Llama+PFT is more capable of accurately estimating the cardinalities of those hard queries, which will be further justified with the end-to-end experiments in Section \ref{sec: end_to_end}. Furthermore, PRICE and LLama + PFT have roughly the same scale of pre-training overhead (6 hours VS 12 hours). This thus indicates that Llama + PFT leads to accurate estimation performance without sacrificing the pre-training efficiency.

\subsection{Main findings}
Our findings from this set of experiments are summarized below:
\begin{itemize}[noitemsep, topsep=0pt, left=0pt]
\item In the workload shift setting, in comparison to the state-of-the-art cardinality estimators, \llm\ exhibit stronger generalizability to the dynamic updates on data and the emergence of new query patterns in the workloads, which can be further enhanced through further fine-tuning.
\item Although fine-tuning raw \llm\ can outperform all baseline methods when data updates happen, the pre-fine-tuned version may suffer from it. This cannot be addressed through further fine-tuning \llm\ on target datasets. To mitigate this, it would be ideal to incorporate both read-intensive and write-intensive workloads into the pre-fine-tuning phase. 

\end{itemize}

%% file: exp_complex_queries.tex
\section{Performance on Complex queries}\label{sec: complex_queries}
In this section, we aim to assess the performance of \llm\ on complex queries, including LIKE queries, DISTINCT queries, and a real-world workload, JOB\cite{JOB}. 

\subsection{Experimental setups}\label{Sec:complex_exp_setup}

\textit{LIKE queries}
For LIKE queries, we follow \cite{aytimur2024lplm} to perform experiments on one dataset, DBLP-AN, which is a single-column table created by extracting the author name field from the DBLP dataset\cite{dblp2019}. 
Since no LIKE queries exist in these datasets, we thus follow \cite{aytimur2024lplm} to randomly generate $n$ string patterns as queries to compose the training and testing set. Those $n$ string patterns are produced by randomly selecting strings from the above single-column datasets and enumerating all possible string patterns until the number of patterns reaches $n$. 
We primarily compare the estimated cardinalities by \llm\ to the state-of-the-art ML-based cardinality estimators for LIKE queries, LPLM \cite{aytimur2024lplm}, which builds a query-driven neural language model to model the probability distribution of general LIKE patterns. Note that LPML requires up to 5 million training queries to reach its best performance.
Since the state-of-the-art data-driven methods such as ASM \cite{kim2024asm} can also support LIKE queries, we additionally compare ASM against \llm\ in this experiment. Same as Section \ref{sec: benchmark}, PG is also included as the baseline.

\textit{DISTINCT queries}
To our knowledge, except \cite{hayek2020nn} which targets DISTINCT queries for \ce\ task but does not release the code, no other solutions exist for dealing with arbitrary DISTINCT queries. To facilitate experiments with DISTINCT queries, the experiments in this subsection concentrate on one subclass of DISTINCT queries in this paper, i.e., the queries with the DISTINCT operation applied to one column. This problem is thus equivalent to estimating the number of distinct values (i.e., NDV) for one column. Considering that the problem of estimating NDV has been extensively studied, we therefore primarily compare \llm\ against one state-of-the-art NDV estimation method, EstNDV \cite{wu2021learning} and PG. In addition, due to the lack of standard \ce\ benchmarks consisting of DISTINCT queries, we produce a synthetic workload by adding a DISTINCT operator to each query of the JOB-light workload \cite{job_light_workload}. 

\textit{JOB workload} The JOB workload is constructed based on the IMDB dataset \cite{leis2015good}, which contains 113 queries in total. Specifically, each query consists of one select-project-join block but with an average of 8 joins and complicated predicates, such as nested logic expressions, BETWEEN and IN operators. Hence, those queries become far more complex than simple Select-Project-Join (SPJ) ones. This workload is usually designed to mimic realistic query patterns, in particular, the patterns with highly correlated joins and predicates. This workload has been widely used to evaluate the capability of cardinality estimators in handling complex queries \cite{kim2024asm}. Query-driven methods and hybrid methods, such as PRICE, only support simple SPJ queries. On the other hand, the state-of-the-art data-driven method, ASM \cite{kim2024asm}, has already demonstrated its superiority on this workload. Therefore, we only compare Llama + FT and Llama + PFT + FT against ASM and PG in this experiment.

\textit{Configurations for \llm}
As revealed in Section \ref{sec: generalizability}, LIKE queries and DISTINCT queries are absent from the pre-training datasets {(noted as PFT)}. {To enhance the capability of \llm\ in handling LIKE and DISTINCT queries, we include such queries into the training sets for the pre-fine-tuning phase (denoted as {$PFT^{Like}$ and $PFT^{Distinct}$}). Specifically, following the query generation procedure in Section \ref{sec: gen_workload}, these LIKE and DISTINCT queries are generated using the same LLM-based SQL query generation method \cite{lao2025sqlbarber}.
}

{To further evaluate the robustness of \llm\ on low-frequency values occurring in the LIKE queries—where coarse-grained statistics (e.g., most frequent values) may not cover—we perform a breakdown analysis on the testing results by distinguishing between high-frequency and low-frequency queries. Specifically, the low-frequency group corresponds to LIKE patterns that do not match any entries in the top-$k$ most frequent value list ($k=5$). Analyzing this group allows us to assess whether the estimators can effectively rely on global statistics (e.g., NDV) and internal knowledge about the query semantics and data distributions  obtained from the pre-fine-tuning and fine-tuning stage.}

\subsection{Experimental results}
The results of experiments with LIKE queries, DISTINCT queries, and the JOB workload are reported in Table \ref{Table: like_accuracy}, Table \ref{Table: distinct_accuracy} and Table \ref{Table: JOB q_error}, respectively. As these tables suggest, for complex queries, \llm\ outperform the baseline methods by a large margin in most cases. \reviewertwo{For instance, \llm\ reduce the 50 percentile Q-error of LPLM by up to 
43.44\% (50 percentile error is reduced from 11.9 to 6.73) as shown in the second column of Table \ref{Table: like_accuracy}. }

{To further examine the robustness of the low-tailed values occurring in the LIKE queries, Table \ref{Table: like_frequency} details the performance on queries using high-frequency values (those belonging to the Top-$K$ most frequent values) versus the remaining low-frequency values. The results demonstrate that \llm\ outperforms baselines in both scenarios. Notably, even for low-frequency queries where corresponding statistics are missing, \llm\ (specifically $PFT^{Like}$) achieves error reduction using its internal knowledge about the query semantics and data distributions obtained from its pre-fine-tuning and fine-tuning stage.}

It is worth noting that in comparison to ML-based estimators, Postgres produces highly inaccurate estimations for both LIKE queries and DISTINCT queries. This thus highlights the necessity of enhancing the \ce\ performance for complex queries beyond simple SPJ queries in practice.

These results also shed light on the capability of \llm\ to handle more general queries. For instance, estimating the cardinalities for group-by queries is the same as that for DISTINCT queries since the number of tuples in the group-by query result is equivalent to the number of unique values occurring in the group-by attributes. Consequently, \llm\ could achieve the same \ce\ accuracy for group-by queries as that for DISTINCT queries. In the future, we will expand this study to investigate the \ce\ performance of \llm\ for arbitrarily complicated queries.

\begin{table}[!h]
\small
    \centering
{ 
\begin{tabular}{c|cccc}
\toprule
\multirow{2}{*}{\ce\ method} & \multicolumn{4}{c}{Q-error} \\
& 50\% & 70\% & 90\%  & {99\%}\\ \hline
PG & 26.5 & 40.5 & 87.5 & {1253.7}\\
LPLM & 11.9 & 17.2 & 36.9 & {432.6} \\
ASM & 22.0 & 36.0 & 53.0 & {862.0} \\
Llama + FT & 8.2 & 11.8 & 24.3 & {318.9}\\
Llama + PFT + FT & 9.11 & 10.9 & 19.37 & {309.2} \\
{Llama + {$PFT^{Like}$} + FT} & {\textbf{6.73}} & {\textbf{9.01}} & {\textbf{15.2}} &{\textbf{235.8}} \\ \hline
\end{tabular}
}
\caption{{Performance on DBLP datasets with LIKE queries}
}\label{Table: like_accuracy}
\end{table}

\begin{table}[!h]
\small
    \centering
    \resizebox{0.5\textwidth}{!}{
\begin{tabular}{c|c|ccccc}\toprule
\multirow{2}{*}{Workloads}& \multirow{2}{*}{\ce\ method}&\multicolumn{4}{c}{Q-error} 
\\
& & 50\% & 70\% & 90\% & 99\% \\\hline
\multirow{6}{*}{{High-Frequency }} & PG & 13.5 & 23.5 & 58.6 & 502.0 \\
& LPLM & 9.8 & 16.4 & 45.0 & 408.1\\
& ASM & 13.0 & 20.0 & 51.0 &  488.0 \\ \hhline{~-----}
& Llama + FT & 7.3 & 10.2 & 12.31  & 249.8\\
& Llama + PFT + FT & 8.15 & 9.36 & 15.43 & 218.5 \\
& {Llama + {$PFT^{Like}$} + FT} & \textbf{5.47} &\textbf{8.75} & \textbf{15.1}  & \textbf{203.79}\\
\hline
\multirow{6}{*}{{Low-Frequency }} & PG & 36.0 & 45.5 & 90.7 & 1287.6 \\
& LPLM & 12.1 & 18.8 & 35.2 & 440.0 \\
& ASM & 23.0 & 37.0 & 54.0 & 870.0 \\ \hhline{~-----}
& Llama + FT & 18.1 & 40.2 & 45.07& 392.7 \\
& Llama + PFT + FT & 13.5 & 16.9 & 35.4 & {379.9} \\
& {Llama + {$PFT^{Like}$} + FT} & \textbf{10.2} & \textbf{15.8} & \textbf{31.6}  & \textbf{378.2}  \\
\bottomrule
\end{tabular}
}
\caption{{Performance of LIKE queries on DBLP: High-Frequency vs. Low-Frequency.}
}\label{Table: like_frequency}
\end{table}

\begin{table}[!h]
\small
    \centering
\begin{tabular}{c|cccc}\toprule
\multirow{2}{*}{\ce\ method}&\multicolumn{4}{c}{Q-error}\\
& 50\%  & 70\% & 90\% & {99\%} \\\hline
 PG & 4.34 & 7.53 & 13.89 & {196.78} \\
EstNDV & 2.15 & 3.63 & 6.93 & {175.43}\\
Llama + FT & 1.87 & 3.35 & 6.74 & {170.21} \\
Llama + PFT + FT & 1.88 & 3.07 & 5.97 & {166.89}\\
{Llama + {$PFT^{Distinct}$} + FT} & {\textbf{1.11}} & {\textbf{1.37}} & {\textbf{4.25}} & {\textbf{161.0}}\\
\hline
\end{tabular}
\caption{{Performance on DISTINCT query workload}
}\label{Table: distinct_accuracy}
\end{table}

\begin{table}[!h]
\small
    \centering
\begin{tabular}{c|cccc}\toprule
\multirow{2}{*}{\ce\ method}&\multicolumn{4}{c}{Q-error}\\
& 50\%  & 70\% & 90\% & 99\%  \\\hline
 PG & 2.87 & 29.89 & 754.33 & 14672 \\
ASM & 4.71 & 66.32 & 2746.91 & 75433 \\
Llama + FT & \textbf{1.88} & 14.78 & 129.94 & 1335.71 \\
Llama + PFT + FT & 2.01 & \textbf{10.63} & \textbf{77.84} & \textbf{1079.65} \\
\hline
\end{tabular}
\caption{Performance on JOB Workload 
}\label{Table: JOB q_error}
\end{table}

\subsection{Main findings}
We summarize our main findings as follows:
\begin{itemize}[noitemsep, topsep=0pt, left=0pt]
    \item When confronting complex queries such as LIKE queries, DISTINCT queries, and the ones occurring in the JOB workload, \llm\ are capable of estimating cardinalities more accurately than the state of the art. 
    \item Among the existing solutions, the state-of-the-art solutions, in particular, the query-driven methods, can either handle simple SPJ queries or some specific classes of complex queries, but not all of them. The results in this section thus justify the potential of \llm\ to serve as cardinality estimators for arbitrary queries.
\end{itemize}

%% file: exp_training_efficiency.tex
\section{Training data efficiency}\label{sec: data_efficiency}
As mentioned in \cite{zeng2024price}, the PRICE model is pre-trained with up to 1.3$\times$10$^6$ queries gathered from 26 diverse datasets. Collecting such a large training dataset is extremely time-consuming in that one has to execute each query to derive the true cardinality. In contrast, data-efficient fine-tuning is an emerging capability of \llm.
Therefore, in this section, we follow \cite{zeng2024price} to explore how the size and diversity of training data used in the pre-fine-tuning phase impact the \ce\ performance of \llm, which is compared against PRICE. 

\subsection{Impact of the number of queries}
To begin with, we perform pre-fine-tuning on \llm\ with varied sizes of pre-training datasets. Specifically, we randomly select $n$ queries from each of the 26 datasets, compose a training dataset of size 26$n$, and vary the value of $n$ from 100 to 9000. To quantify the impact of $n$, we evaluate the generalizability of the pre-fine-tuned \llm\ to unseen datasets (without additional fine-tuning), for which we utilize the four benchmark datasets from Section \ref{sec: benchmark}. To determine an appropriate value of $n$, we also evaluate the \ce\ performance of \llm\ on a held-out query set from the 26 pre-training datasets. Note that during the pre-fine-tuning phase, the estimates of PRICE are excluded from the prompt for \llm. 

\begin{figure*}
    \centering
    \includegraphics[width=0.85\linewidth]{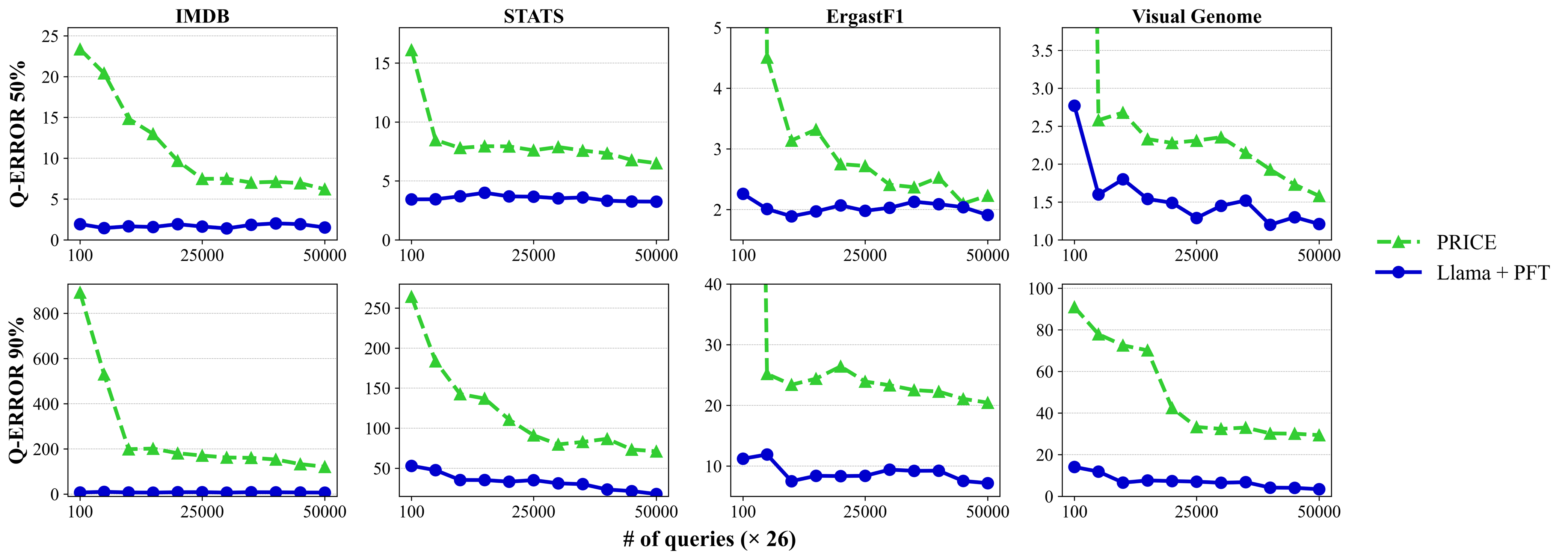}
    \caption{\ce\ performance of pre-fine-tuned Llama and pre-trained PRICE with varied numbers of queries}
    \label{fig:num_queries}
\end{figure*}

The experimental results are visualized in Figure \ref{fig:num_queries}, which suggests that the pre-fine-tuned \llm\ constantly outperform the pre-trained PRICE model. In addition, with smaller pre-training dataset sizes, the performance gains of Llama+PFT gradually grow, thus indicating that it might be unnecessary to prepare large amounts of pre-training queries for pre-fine-tuning \llm. 

\subsection{Impact of the number of pre-training datasets}
\begin{figure*}
    \centering
    \includegraphics[width=0.85\linewidth]{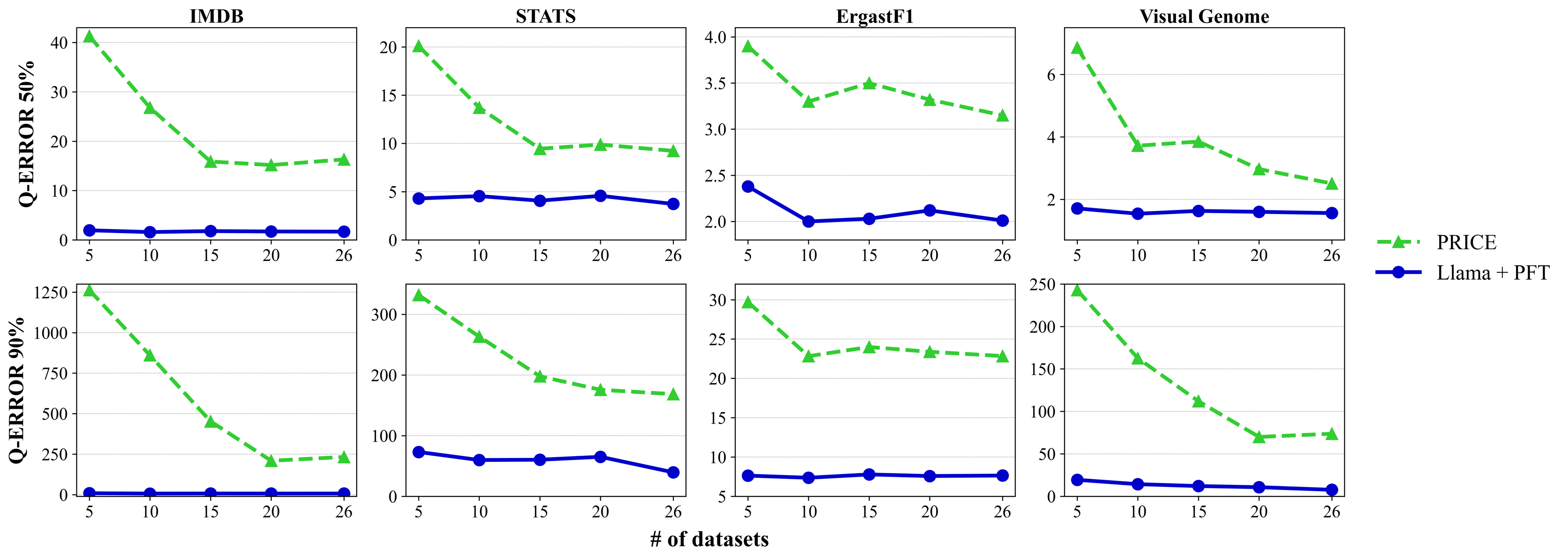}
    \caption{\ce\ performance of pre-fine-tuned Llama and pre-trained PRICE with varied numbers of pre-training datasets}
    \label{fig:num_dataset}
\end{figure*}

As discussed in \cite{zeng2024price}, the number of pre-training datasets reflects the diversity of training data and can also influence the \ce\ performance. We therefore compare the \ce\ performance of Llama and PRICE with varied numbers of pre-training datasets, ranging from 1 to 26. For each dataset, the number of queries is configured as 8000 for both Llama and PRICE for fair comparison. We collect both 50-percentile and 90-percentile Q-errors for Llama and PRICE and depict them in Figure \ref{fig:num_dataset}. As this figure suggests, Llama outperforms PRICE in almost all cases. Also, the Q-errors of Llama only experience slight fluctuation across varied numbers of datasets. This indicates that it might be sufficient to only use a few datasets to pre-fine-tune \llm, thus further saving the data collection efforts. 

%% file: ablation.tex
\section{Ablation study}\label{sec: ablation}
In this section, we first conduct ablation studies on the effect of various \llm\ on the \ce\ task, which is additionally compared against the conventional query-driven \ce\ paradigm by regarding \llm\ as a query encoder and adding one additional MLP layer to predict cardinalities as numbers (denoted by \llm\ + MLP). These results are reported in Table \ref{Table: llm_impact}. \reviewerone{Additionally, we empirically evaluate \llm\ of varying scales within the same family to validate the impact of model scale on \ce\ performance. The results are presented in Table \ref{Table: scale}.} In what follows, we assess the impact of 
critical components of our methods, including few-shot in-context learning, self-correction mechanism, and other estimator information, on \ce\ performance, which are reported in Table \ref{Table: ablation}.

\textit{Impact of different \llm.}
As mentioned in Section \ref{sec: exp_setups}, GPT-4o is much larger than Llama and DeepSeek. Table \ref{Table: llm_impact} can thus shed light on the impact of the model sizes of \llm\ on \ce\ performance. Generally speaking, with the same configuration, say fine-tuning the raw version of \llm, GPT-4o produces smaller Q-errors than the other two on \imdb\ dataset. Despite this, the Q-error reduction brought by GPT-4o is only around 10\%-20\% while fine-tuning it incurs much higher computational overhead ($\sim$7 hours for GPT-4o VS $\sim$2 hours for Llama). By comparing Llama and DeepSeek, their performance numbers are roughly the same since their model sizes are very close (8B VS 16B). 

In addition, as the first line of Table \ref{Table: llm_impact} shows, the Q-errors of Llama + MLP are extremely huge (over 200). This thus invalidates the conventional modeling paradigm that employs \llm\ as encoders for \ce\ task. To further understand why Llama + MLP is much less effective in comparison to our proposed strategy, we analyzed the following randomly selected query $Q$ from the test split of JOB-light workloads: ``{\it SELECT COUNT(*) FROM movie\_keyword imdb\_mk, cast\_info imdb\_ci WHERE (imdb\_mk.movie\_id = imdb\_ci.movie\_id) AND (imdb\_ci.role\_id = 6)}'' alongside the training query that is most similar to this test query. Such similar training queries are discovered in two different ways, denoted by $Q1$ and $Q2$, which leverage the embeddings of Llama + MLP and Llama + FT respectively. The analysis reveals that the ground-truth cardinality of $Q$ is significantly closer to that of $Q2$ than $Q1$. This finding suggests that in comparison to \llm\ + MLP, the query embeddings produced by \llm\ + FT can better align with the query cardinalities.

\begin{table}[!h]
\small
    \centering
\begin{tabular}{c|cccc}\toprule
\multirow{2}{*}{\ce\ method}&\multicolumn{4}{c}{Q-error}
\\
& 50\% &90\% & 95\% & 99\% \\\hline
LLama + MLP &298&1774&3688&93471\\ \midrule
Llama + FT &1.64&9.26&23.36&35.85\\
DeepSeek + FT &1.92&6.04&\textbf{12.95}&208.73\\
GPT-4o + FT &\textbf{1.49}&\textbf{5.39}&19.52&\textbf{26.79}\\\midrule
Llama + PFT &1.65&6.24&\textbf{12.68}&\textbf{26.53}\\
DeepSeek + PFT &\textbf{1.64}& \textbf{6.08} & 34.99 & 807.89 \\ \midrule
Llama + PFT + FT &1.60&\textbf{5.42}&\textbf{7.15}&\textbf{21.06}\\
DeepSeek + PFT + FT &\textbf{1.37}& 5.97 & 14.23 & 89.82\\
\bottomrule
\end{tabular}
\caption{Impact of different \llm\ on \imdb\ dataset}\label{Table: llm_impact}
\end{table}

{
\textit{Impact of \llm\ model scale.}
To provide a more thorough comparison of the impact of model scale on \ce\ performance, we conduct experiments on \llm\ from the same family, Qwen3\cite{yang2025qwen3}, with varying parameter scales. We utilize Qwen3-0.6B, 1.7B, 4B, and 8B models and perform PFT, FT, and PFT+FT experiments on each. All FT procedures are carried out on the \stats\ dataset.
The results in Table \ref{Table: scale} show that as model size increases, the performance of PFT, FT, and PFT+FT consistently improves, eventually surpassing that of PRICE. This demonstrates that larger LLMs achieve stronger performance, mainly because larger-scale \llm\ possess greater semantic understanding and generalization abilities, allowing them to capture complex SQL query semantics and underlying data distributions more effectively. Moreover, we test the Qwen3-8B base model directly on \ce\ task without any fine-tuning, and observe that while it already exhibits a certain level of \ce\ capability, its performance is inferior to PRICE and significantly behind the fine-tuned models. This confirms that task-specific fine-tuning can substantially enhance the specialized capabilities of \llm.
}

\begin{table*}[h!]
\centering
\begin{tabular}{c|c|c|c|c}
\toprule
 & PFT & FT & PFT+FT & Base \\
 & 50th / 90th / 99th & 50th / 90th / 99th & 50th / 90th / 99th & 50th / 90th / 99th \\ \hline
Qwen3-0.6B & \multicolumn{1}{l|}{5.39 / 125.91 /{21860.26}} & 1.49 / 7.12 / 80.15 & 1.42 / 6.62 / 75.83 & - \\
Qwen3-1.7B & 5 / 120 / {105501.23} & 1.47 / 5.27 / 50.46 & 1.31 / 4.65 / 50 & - \\
Qwen3-4B & 4.76 / 105.05 / {11958.50} & 1.28 / 4.71 / 38.9 & 1.17 / 4.47 / 37.51 & - \\
Qwen3-8B & \textbf{4.35} / \textbf{97.76} / \textbf{{8604.53}} & \textbf{1.22} / \textbf{4.43} / \textbf{22.93} & \textbf{1.09} / \textbf{4.15} / \textbf{22.26} & 5.35 / 216.87 / {25078.84} \\ \hline
PRICE & 4.98 / 110 / {9654.00} & - & 2.91 / 33.31 / 697 & - \\ 
\bottomrule
\end{tabular}
\caption{\protect\reviewerone{Impact of \llm\ model scale on Q-Error over the STATS dataset}}\label{Table: scale}
\end{table*}

\begin{figure}[h]
    \centering
    \includegraphics[width=0.9\linewidth]{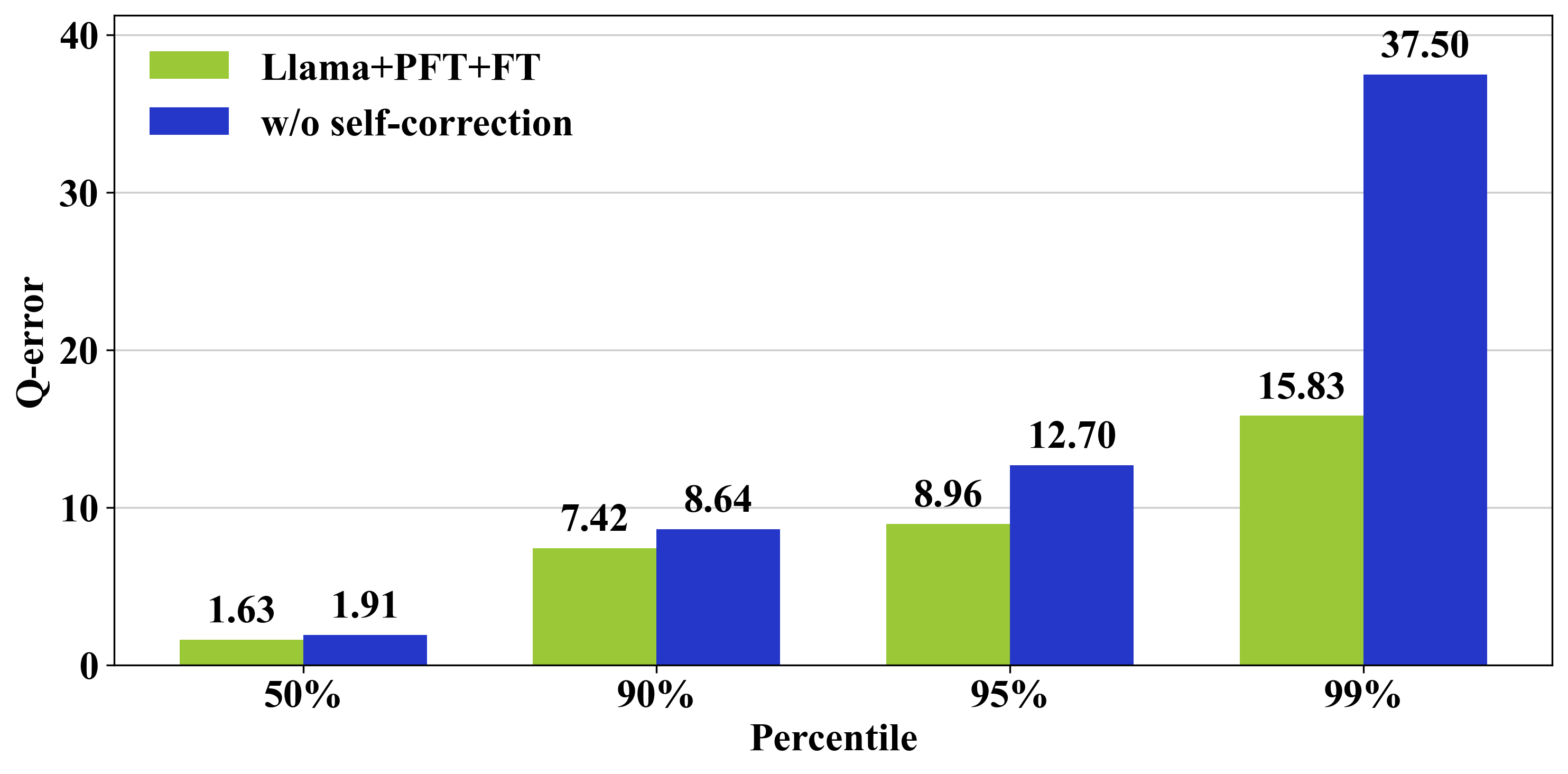}
\caption{\reviewertwo{Ablation study under OOD settings on the JOB-Light dataset}}
    \label{fig:ablation_ood}
\end{figure}

\textit{Impact of self-correction mechanism.}
We evaluate the \ce\ performance of \llm\ with and without the self-correction mechanism. According to Table \ref{Table: ablation}, without this mechanism, the Q-errors are increased across almost all percentiles, which are by up to 144\% (see 99 percentile column). Hence, this justifies the efficacy of the self-correction mechanism in mitigating estimation errors. 
\reviewertwo{To further assess the robustness of self-correction, we conduct an additional ablation study under OOD conditions. Following the setup in Section~\ref{sec: gen_workload}, the models are fine-tuned on the CEB dataset and evaluated on the JOB-light workload. As shown in Figure~\ref{fig:ablation_ood}, the comparison between Llama+PFT+FT and its variant without self-correction reveals that the self-correction mechanism consistently improves Q-error performance across all percentiles. These results confirm that self-correction effectively reduces the Q-errors under workload-shift scenarios.}

\textit{Impact of coarse-grained statistics.}
Although coarse-grained statistics cannot fully capture the entire data distribution, it is still essential to incorporate them in prompts. Otherwise, as suggested by Table \ref{Table: ablation}, the \ce\ performance of \llm\ drops by a large magnitude, which is up to 254\% in 99-percentile Q-error. 

\textit{Impact of other estimator information.} 
We also empirically assess the impact of other estimator information, i.e., the estimated cardinalities from other \ce\ methods. As introduced in Section \ref{sec: method}, this information is also required for the self-correction mechanism, which is disabled in this experiment to present its interference. The experimental results in Table \ref{Table: ablation} indicate that after excluding other estimator information from prompts, the \ce\ performance experiences at least one order-of-magnitude degradation, thereby demonstrating the critical role of such information. 

We also compared the results of only using the estimated information from PG (see the row ``w/ only PG information''). This means that the estimated results by PRICE are excluded from the prompts of \llm\ and the PG estimates are employed for the self-correction process instead. Table \ref{Table: ablation} suggests that the \ce\ performance of this configuration is worse than the default configuration of Llama + PFT + FT, thus highlighting the necessity of incorporating an accurate cardinality estimator in the prompts and the self-correction process for \llm.

\textit{Impact of few-shot examples.} 
As introduced in Section \ref{sec: method}, we do not use few-shot examples in the prompts for \llm\ by default. In this experiment, we aim to assess whether adding few-shot examples could bring additional performance gains or not. We thus randomly select 20 queries with their ground-truth cardinalities as few-shot examples and append them to the prompt for Llama+PFT+FT. 

As Table \ref{Table: ablation} suggests, adding few-shot samples in prompts surprisingly hurts \ce\ performance. 
Therefore, further studies, say on appropriate strategies for selecting few-shot samples, would be essential in the future.

\begin{table}[!t]
\small
    \centering
    \resizebox{0.5\textwidth}{!}{
\begin{tabular}{l|cccc}\toprule
\multirow{1}{*}{\ce\ method}& \multicolumn{4}{c}{Q-error}
\\
& 50\% & 90\% & 95\% & 99\% \\\hline
Llama+PFT+FT & \textbf{1.33} & \textbf{5.77} & 9.57 & \textbf{30.90}\\
\hspace{10pt} w/o self-correction & 1.36 & 6.38 & \textbf{8.31} & 59.55 \\
\hspace{20pt} w/o other estimator(include PG) information & 171.12 & 10999 & 16434 & 37281\\
\hspace{10pt} w/ only PG information & 2.19 & 11.82 & 32.38 & 69.64\\
\hspace{10pt} w/o coarse-grained statistics & 1.34 & 5.85 & 14.11 & 86.19\\
\hspace{10pt} w/ few-shot examples & 1.77 & 11.75 & 59.84 & 397.21 \\
\hspace{10pt} {w/  schema anonymization (inference only)} & {1.69} & {6.47} & {20.29} &{56.32}  \\
\hspace{10pt} {w/o bootstrap resampling} & {1.59} & {6.39} & {10.23} & {34.28}\\\hline
{PRICE}&{2.91} & {33.31}&{69.75} &{697}\\
\hspace{10pt} {w/o bootstrap resampling}& {3.18} & {34.11} &{69.75} &{697}\\
\hspace{10pt} {w/  schema anonymization (inference only)} & {4.17} & {29.06} & {69.73} &{2354} \\\hline
{ALECE}& {1.67} & {7.93}&{18.56} &{119}\\
\hspace{10pt} {w/o bootstrap resampling}& {1.67} & {7.93} &{18.57} &{119}\\
\bottomrule
\end{tabular}
}
\caption{\reviewerone{Ablation study on \stats\ dataset }
}\label{Table: ablation}
\end{table}

{
 \textit{Impact of dataset contamination mitigation.}
 To evaluate the potential impact of dataset contamination on the \stats\ dataset, we anonymize all table and column names (e.g.,  “tableA.colB”). The results in Table \ref{Table: ablation} (see the row “w/ schema anonymization”) show comparable accuracy to the original setup (see the row "Llama+PFT+FT"). This demonstrates that our model almost does not rely on memorized schema information from publicly available datasets such as STATS.
 We additionally report the \ce\ performance of PRICE in the same setting in Table \ref{Table: ablation}, which shows that the Q-error of PRICE at the 50th percentile is increased by around 40\%.
This thus demonstrates that our method almost does not rely on memorized knowledge about the STATS dataset and learns more generalizable representations for query semantics and data distribution than PRICE.
 }

 \textit{Impact of bootstrap resampling.}
 To evaluate the effect of bootstrap resampling on the baseline method PRICE, we follow the strategy described in Section~\ref{sec: glm_prompt}. Specifically, multiple PRICE models are trained on bootstrapped versions of the training set. For each query, the final prediction is selected from the model that exhibits the highest confidence, measured by the narrowest bootstrap interval. 
\revision{As summarized in Table~\ref{Table: ablation} (see the row “PRICE w/o bootstrap resampling”), omitting bootstrap resampling leads to slightly higher Q-errors compared with the version with resampling, especially in the 50th and 90th percentiles. These performance degradations indicate that without the variance reduction provided by resampling, the estimation accuracy of PRICE is slightly compromised due to increased instability.}

 \revision{Given that the estimates provided by PRICE serve as critical contextual information for the LLMs, their reliability directly influences the downstream generation quality. Consequently, omitting the bootstrap resampling step exposes the LLM to potentially noisy and unstable estimates from PRICE, thereby degrading the overall performance. This is evident in the row “Llama+PFT+FT w/o bootstrap resampling”, where the exclusion of the resampling module results in even higher Q-errors, demonstrating that the robustness of the prompt information is essential for maximizing the LLM's reasoning capabilities.}

\revision{
To further justify our design of applying bootstrap resampling to PRICE, we evaluated the impact of this strategy on other strong baselines, specifically ALECE. As reported in Table \ref{Table: ablation} (see the row ``ALECE'' and ``ALECE w/o bootstrap resampling''), incorporating bootstrap resampling into ALECE yields negligible performance improvements compared to its standard version. This suggests that while bootstrap resampling is critical for reducing variance in the learning-based PRICE model, it offers marginal accuracy gains for baselines like ALECE, which are inherently more deterministic. Therefore, to balance accuracy and efficiency, we employ this strategy within the PRICE module.
}

%% file: exp_end2end_time.tex
\section{End-to-end evaluation}\label{sec: end_to_end}

In this section, we adopt the configuration of \cite{zeng2024price} to inject the estimated cardinalities of each \ce\ method into Postgres using PilotScope \cite{zhu2024pilotscope}. This setup allows us to measure the end-to-end query execution time (E2E time) in the DBMS.  
To measure the E2E time for one query, we follow \cite{zeng2024price} to first decompose it into smaller sub-queries. The estimated cardinalities for these sub-queries are then input into the DBMS query optimizer to construct query plans. The \revision{query execution} time is based on these constructed query plans. This evaluation is performed on both the SPJ query workloads from Section \ref{sec: benchmark} and the JOB workload \cite{JOB}, one realistic workload based on the IMDB dataset. 
The decomposed sub-queries for these two types of workloads are sourced from \cite{zeng2024price} and \cite{negi2021flow} respectively.

\reviewertwo{\subsection{Pre-fine-tuned models}}\label{sec:e2e_pft}
\reviewertwo{\subsubsection{Experimental setups}}

To emulate the real-world scenario, we first consider a scenario where fine-tuning \llm\ and baseline models is infeasible due to hardware constraints. Hence, we compare Llama + PFT against the baseline methods including PG, the pre-trained PRICE model (i.e., PRICE (w/o FT)) and data-driven methods. Prior studies \cite{zeng2024price,kim2024asm} have shown that PRICE (w/o FT) and ASM outperform mainstream data-driven methods including NeuroCard \cite{yang2020neurocard}, DeepDB \cite{hilprecht13deepdb} and FactorJoin \cite{wu2023factorjoin} in most E2E experiments. Therefore, we only compare pre-fine-tuned Llama, i.e., Llama+PFT against PG, ASM and PRICE (w/o FT) in this experiment. In our evaluation within the JOB workload, a direct comparison with PRICE was not conducted due to its inherent limitations in handling complex queries, as PRICE is specifically designed to support only simple SPJ (Select-Project-Join) queries.

\reviewerthree{\subsubsection{Selective use of LLMs}\label{sec: selective_use_llm}}
{Given the extremely large model sizes of \llm, \\ Llama+PFT can introduce significant inference overhead, thus slowing down the overall query execution process, in particular for those queries with short running time. To mitigate this issue, we propose to selectively use Llama+PFT to estimate cardinalities for those potentially expensive sub-queries while employing PG for the remaining ones---a strategy termed \llm\ + PFT + cost. \reviewerthree{These two types of queries are referred to as high-cost queries and low-cost queries, respectively.} To identify which queries are high-cost ones or not, we utilize the Postgres cost model to estimate the execution cost of each sub-query first and then flag those sub-queries with estimated cost over some threshold as potentially expensive. 
This threshold is determined by empirically analyzing the relationship between the estimated cost and the actual running time for a small set of queries, selecting the cost number where the running time starts to rise sharply. Empirical evidence in Section \ref{sec: e2e_pft_result} supports the effectiveness of this strategy. \reviewerthree{This design is also applicable to fine-tuned LLMs.}
}

\begin{figure}[h]
    \centering
    \includegraphics[width=0.9\linewidth]{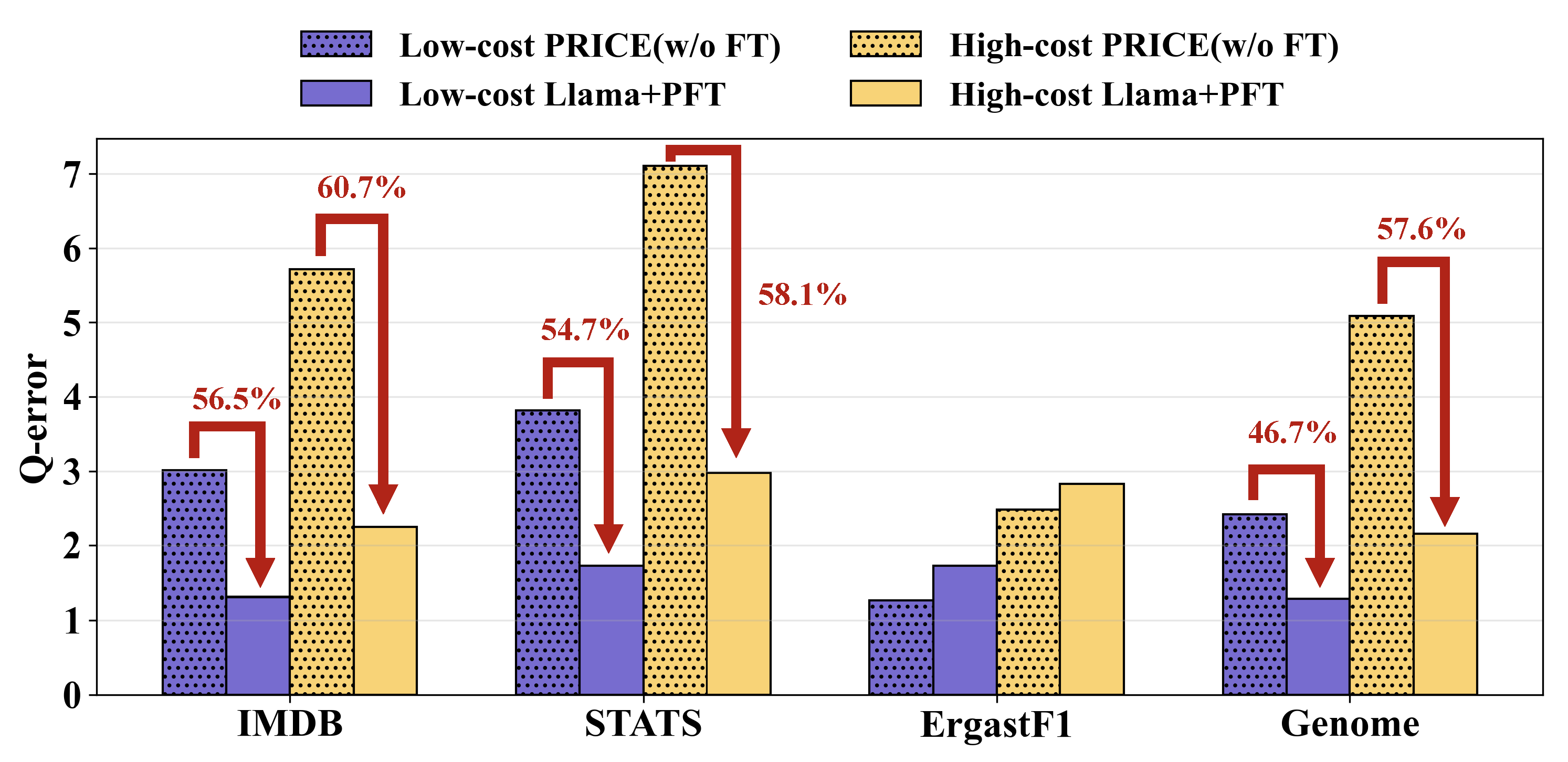}
    \caption{Q-errors at 50th percentile evaluated on low-cost queries and high-cost queries}
    \label{fig:selective_use_llm}
\end{figure}

\reviewerthree{To better justify the rationality of this design, we further evaluate PRICE and our method on both low-cost and high-cost queries, as shown in Figure \ref{fig:selective_use_llm}. The results show that for low-cost queries, existing baseline methods can already provide reasonably accurate estimates. However, for high-cost queries, their estimation accuracy degrades significantly, whereas our method achieves much better performance. Therefore, employing the LLM selectively for high-cost queries yields much more 
performance gains than that for low-cost ones.
The only exception is the ErgastF1 dataset, where our method underperforms PRICE, which is consistent with the results in Table \ref{Table: over_accuracy}. }

\begin{figure}[t]
    \centering
    \includegraphics[width=0.9\linewidth]{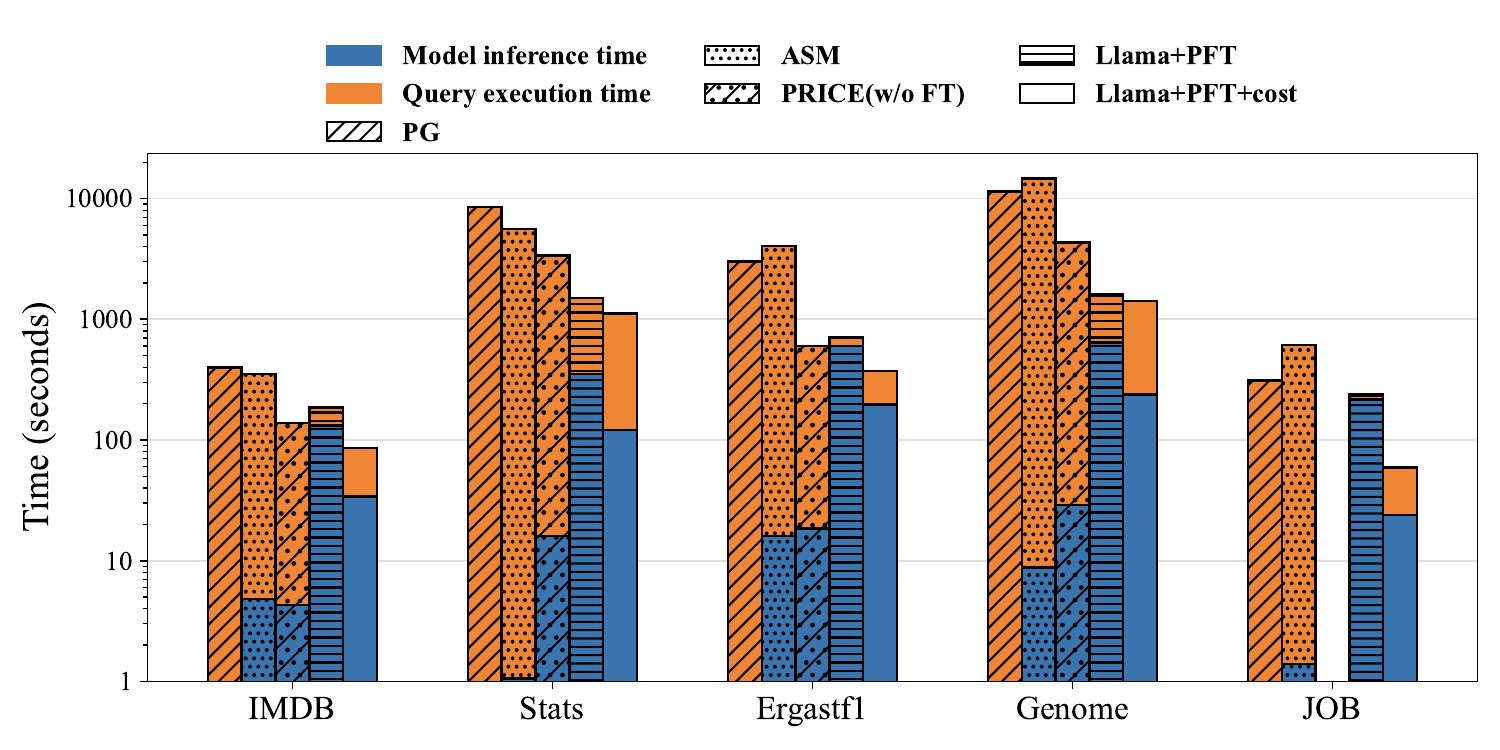}
    \caption{{E2E} time performance with pre-fine-tuned models}
    \label{fig:e2e}
\end{figure}

\reviewertwo{\subsubsection{Experimental results}\label{sec: e2e_pft_result}}
The results are reported in Figure \ref{fig:e2e}, where the \revision{E2E} time represents the accumulated model inference time and \revision{query execution} time across all SQL queries within each workload. \reviewerfour{For LLM-based methods, the model inference time already includes the inference time of the underlying baseline estimators used in the prompts and the time spent on obtaining pre-computed coarse-grained statistics, }\reviewertwo{which accounts for about 2.3\% of the total LLM inference time. } 

Notably, this figure suggests that Llama+PFT reduces the \revision{query execution} time of baseline methods including PG, PRICE and ASM, by a large margin except on \ergast\ dataset, which
is only slightly worse than optimal. This thus highlights the remarkable capabilities of \llm\ for query optimizations without requiring any additional fine-tuning. We also report the model inference time of the pre-fine-tuned Llama and pre-trained PRICE alongside their \revision{query execution} time in Figure \ref{fig:e2e}. After taking the model inference overhead into account, Llama+PFT achieves a shorter overall running time than the baseline methods on \stats\ and \vg\ dataset. However, its overall running time exceeds that of PRICE on JOB-light and \ergast. {In contrast, Llama + PFT + cost only needs to leverage Llama to estimate cardinalities for around 25\%-35\% of sub-queries, thus significantly reducing model inference time by approximately 66\% across all workloads. This strategy results in only a marginal increase in the \revision{query execution} time, making its overall \revision{E2E} time (\revision{query execution} time + model inference time) shorter than that of all baseline methods on all workloads. These results thus demonstrate the potential of leveraging Llama + PFT + cost in real-world DBMSs.}

\begin{figure}[h]
    \centering
    \includegraphics[width=0.9\linewidth]{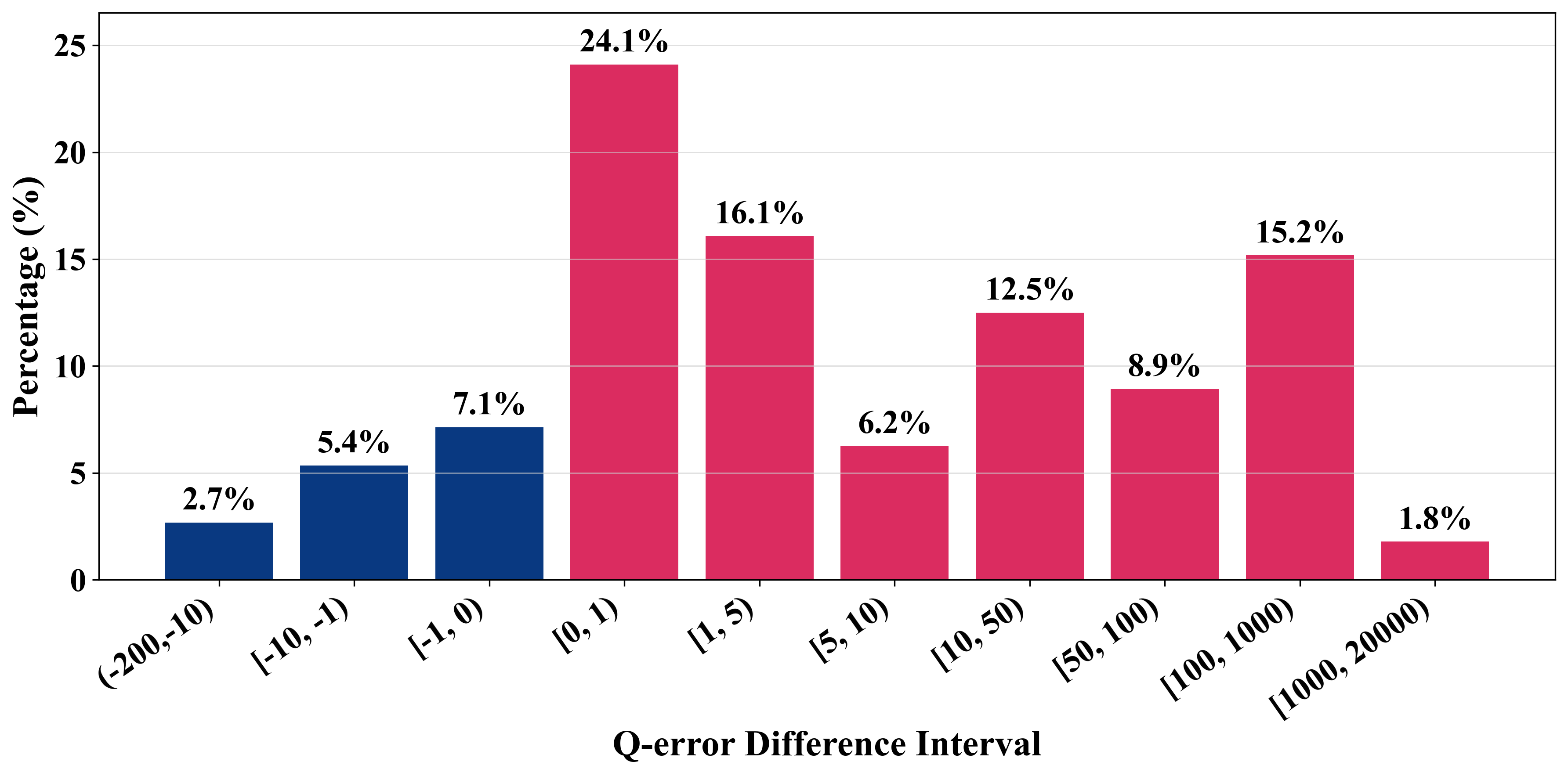}
    \caption{\reviewerthree{Distribution of Q-error reduction after applying self-correction on the \imdb\ dataset}}
    \label{fig:qerror_distribution}
\end{figure}

\begin{figure}[h]
    \centering
    \includegraphics[width=0.9\linewidth]{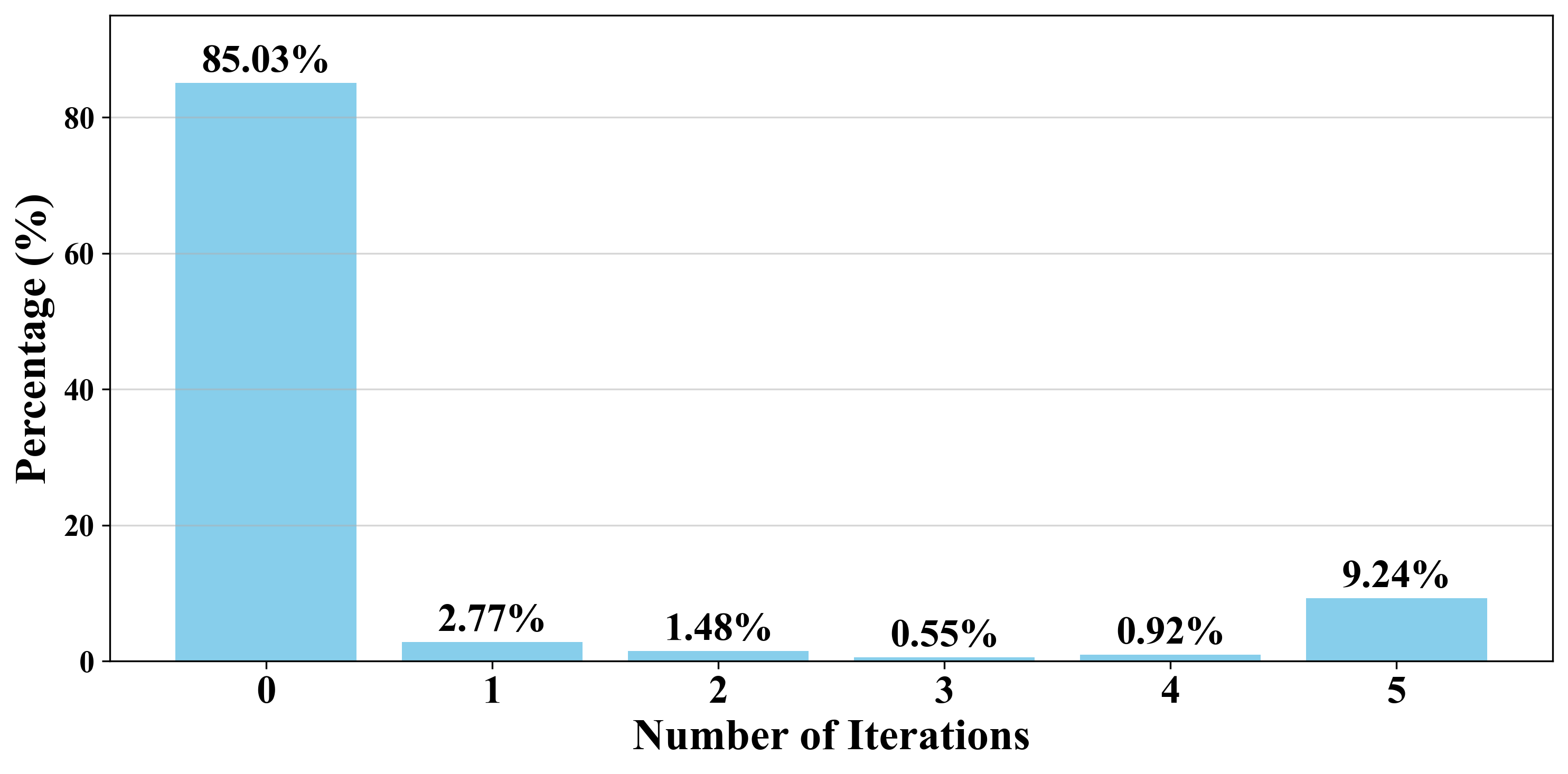}
    \caption{\reviewertwo{Distribution of number of self-correction iterations on \imdb\ dataset}}
    \label{fig:num_iterations}
\end{figure}

\begin{figure}[h]
    \centering
    \includegraphics[width=0.9\linewidth]{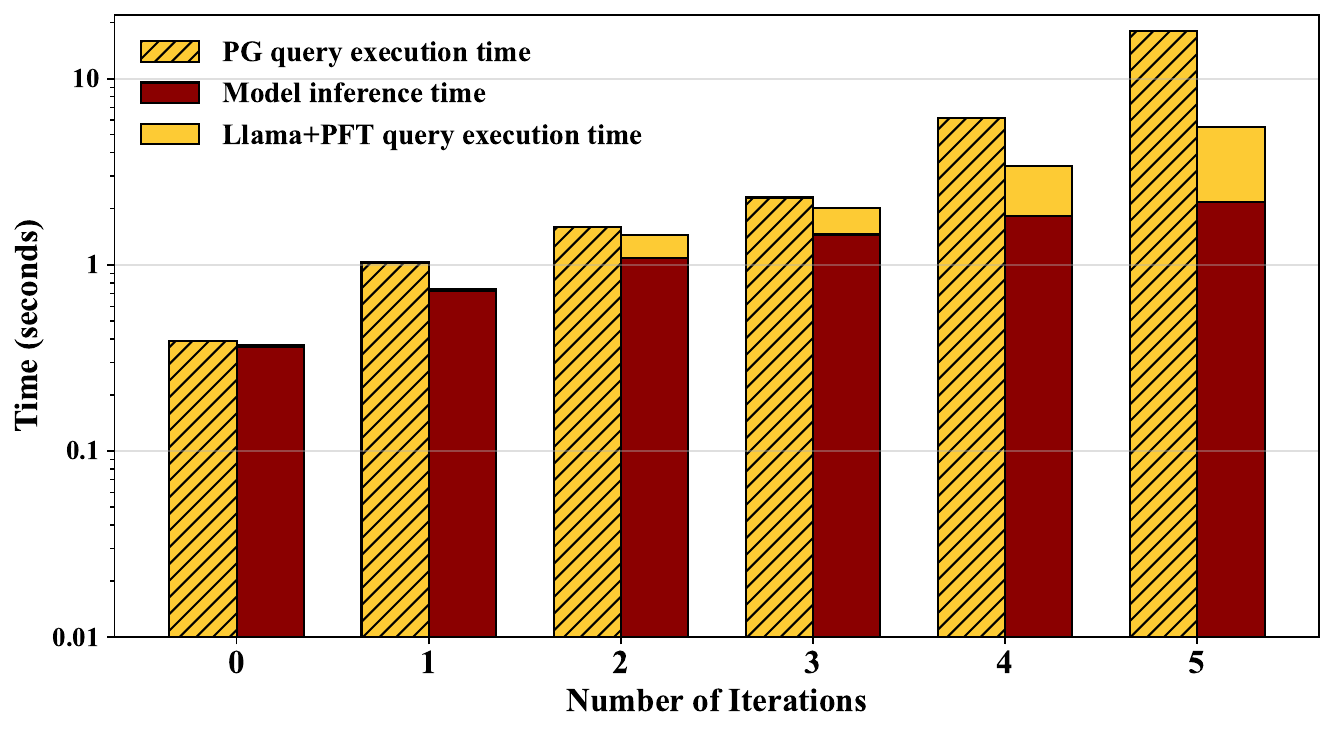}
    \caption{\reviewertwo{Average end-to-end execution time of queries with varying numbers of self-correction iterations on the \imdb\ dataset}}
    \label{fig:iter_e2e}
\end{figure}

{In the inference process of LLM-based methods, we employ a self-correction mechanism to obtain more accurate cardinality estimates. }{ As shown in Figure~\ref{fig:qerror_distribution}, we report the distribution of Q-error reductions achieved by the Llama+PFT model after applying self-correction on the IMDB dataset. The results indicate that while a small fraction of queries experience minor degradation, the majority benefit from self-correction, which generally leads to consistent improvements in estimation accuracy. }\reviewertwo{
Regarding the computational cost introduced by self-correction, Figure~\ref{fig:num_iterations} presents the distribution of the number of self-correction iterations required for each query on the IMDB dataset. We observe that 85.03\% of queries can achieve sufficiently accurate estimates without any self-correction, while 9.24\% of queries require up to five iterations to converge to satisfactory results.
Furthermore, Figure~\ref{fig:iter_e2e} compares the average query execution time and model inference time between different numbers of self-correction iterations. Although queries with more iterations incur a longer model inference time, they also achieve a more significant reduction in the query execution time, thus effectively decreasing the overall end-to-end latency. These results suggest that the additional cost of self-correction is well justified by the performance gains it brings.
}

{\subsection{Fine-tuned models}\label{sec: e2e_ft}}

{\subsubsection{Experimental setups}}\label{sec: e2e_ft_setups}

To further investigate the potential performance gain brought by model specialization, we also consider fine-tuning LLMs and baseline models on task-specific datasets. The fine-tuned Llama model, denoted as Llama+PFT+FT, is expected to capture workload-specific characteristics and improve both estimation accuracy and execution efficiency.
For comparison, we also fine-tune the outstanding baseline methods, including PG and PRICE\cite{zeng2024price}
using the same training corpus to ensure fairness. \revision{Specifically, for PRICE, we start with the original pre-trained checkpoint provided by the authors and perform further fine-tuning on our target corpus.} The design of Llama+PFT+FT+cost follows the same setting described in Section \ref{sec: selective_use_llm} and PRICE+PG is a selective estimation variant that employs PG for low-cost queries and PRICE for high-cost queries.

\begin{figure}
    \centering
    \includegraphics[width=0.9\linewidth]{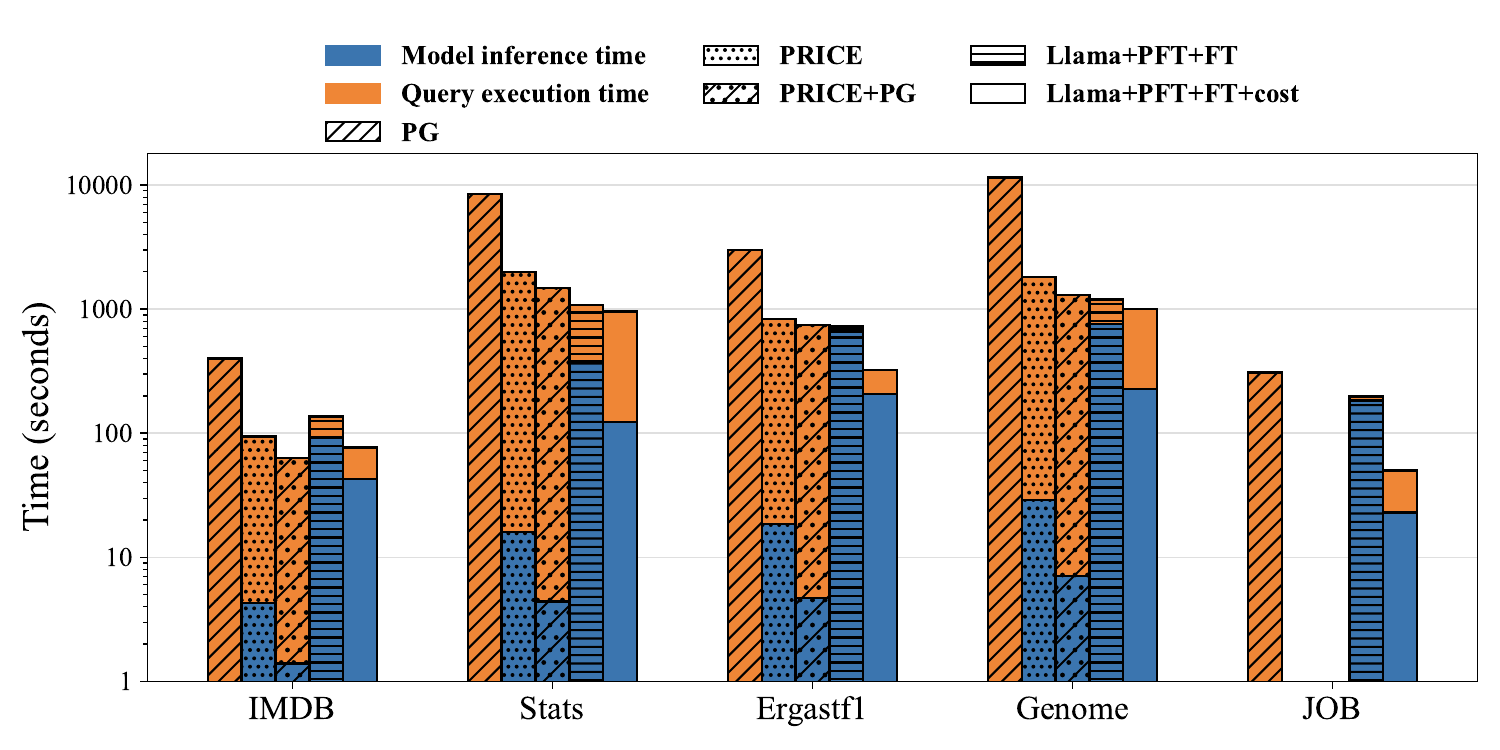}
    \caption{{E2E} time performance with fine-tuned models}
    \label{fig:e2e_ft}
\end{figure}

{\subsubsection{Experimental results}\label{sec: e2e_ft_result}}
The results are reported in Figure \ref{fig:e2e_ft}, where the meanings of model inference time and \revision{query execution} time are consistent with those defined in Section \ref{sec: e2e_pft_result}. As shown in the figure, Llama+PFT+FT further reduces the \revision{query execution} time compared with baseline methods except on the IMDB dataset. This improvement mainly stems from their enhanced estimation accuracy, which leads to more efficient query execution plans. Even when accounting for the relatively long model inference time, Llama+PFT+FT still demonstrates a reduction in total time by up to 33\% on the STATS, ErgastF1, and Genome datasets.

{In addition, Llama+PFT+FT+cost only invokes the LLM for a subset of sub-queries. While this selective strategy incurs a slight sacrifice in estimation accuracy, it significantly reduces model inference time. Llama+PFT+FT+Cost outperforms PRICE on the IMDB dataset, and also further amplifies the total time reduction on the STATS, ErgastF1, and Genome datasets, demonstrating the effectiveness of cost-aware selective estimation in balancing efficiency and accuracy.}

We also observe that the trend of PRICE+PG in Figure \ref{fig:e2e_ft} is consistent with the accuracy results in Table \ref{Table: over_accuracy}—that is, while PRICE+PG achieves shorter running time than PRICE and Llama+PFT+FT on the IMDB dataset due to reduced inference time, it still does not surpass the Llama+PFT+FT on other datasets.

\revision{Specifically, the advantage of PRICE+PG on the IMDB workload is primarily due to the significantly lower model inference latency of PRICE compared to the \llm. While the \llm\ achieve a competitive query execution time on this workload, its cumulative inference overhead becomes more prominent here than in other datasets. However, on more complex workloads like STATS and Genome, the superior estimation accuracy of the LLM leads to execution plan improvements that far outweigh its inference cost, resulting in the best overall performance.}

In addition, except in the IMDB dataset, where PRICE+PG achieves comparable overall overhead in comparison to Llama+PFT+FT+Cost, Llama+PFT+FT+Cost outperforms PRICE and PRICE+PG by a large margin (up to 56\%) on other datasets. This thus demonstrates the significant potential of this strategy in providing accurate estimations while enabling efficient model inference.

%% file: discussion.tex
\section{Discussion}\label{sec: discussion}
\subsection{Lessons learned}
\textit{Can the generalized language models (\llm) be a candidate cardinality estimator for DBMS?} 
Our extensive investigations have revealed that careful prompting, fine-tuning, and inference strategies enable LLMs to generate numeric tokens for database query cardinalities, which often outperform state-of-the-art ML-based estimators. Unlike conventional ML-based \ce\ methods, which are limited to specific settings, LLMs demonstrate versatility and state-of-the-art performance across diverse scenarios with minimal prompt adjustments.


\textit{Are \llm\ ready to serve for cardinality estimation in practice?} Although \llm\ have shown their potential for \ce\ task, there are still some challenges to be addressed before they can be integrated into DBMS. One notable issue is the huge inference overhead of \llm, which can slow down the overall query optimization process.
Therefore, one potential future step toward the use of one LLM as a practical cardinality estimator is to accelerate its inference speed while maintaining its \ce\ performance. This paper takes an initial step in this direction by \emph{selectively} estimating potentially costly queries using \llm. The results demonstrate the practical potential of employing \llm\ in CardEst.

\textit{The role of pre-fine-tuning.} Pre-fine-tuning plays a key role in the \ce\ performance of \llm. However, in this paper, we reuse the pre-training datasets from \cite{zeng2024price} for pre-fine-tuning, which may not be optimal for all scenarios. For example, as revealed in Section \ref{sec: gen_data_dist}, the pre-fine-tuned \llm\ perform worse than the raw \llm\ when dealing with write-intensive workloads since the pre-training datasets are all composed of queries from read-only workloads during the pre-fine-tuning phase.

\textit{Will employing \llm\ for \ce\ incur significant monetary cost?}
\reviewerone{One of the biggest concerns of using \llm\ is its inevitable monetary cost. However, our approach assumes local deployment, where inference runs on a single NVIDIA RTX 4090 GPU. In this setting, the marginal cost is minimal, with no additional cost beyond the hardware itself. In our experiments, all \llm\ were deployed locally on a single GPU without any API calls, incurring negligible additional cost while still achieving surprisingly strong performance on cardinality estimation.}

\reviewerone{Moreover, we observe that the deployment of our methods for large-scale cloud database services remains cost-effective. Specifically, we have examined the pricing of commercial cloud database services and found that the cost of renting and maintaining large-scale cloud databases is typically much higher than that of renting a single GPU for local inference. Therefore, when the database size grows large, the inference cost of our LLM-based estimator becomes relatively insignificant compared to the overall operational cost of the database system. For instance, as suggested by AWS\footnote{see pricing details at \url{https://aws.amazon.com/rds/pricing/?utm_source=chatgpt.com}}, it would take around 2,700 dollars each month to host a Postgres instance for a 1TB database, whereas renting a 24GB NVIDIA RTX 4090 GPU to host the fine-tuned LLMs would only take around 300 dollars\footnote{see pricing details at \url{https://www.cloudrift.ai/pricing?utm_source=chatgpt.com}}. This means that we only add around 10\% more monetary cost to deploy our solution to a real cloud database, which is a feasible option.}

\subsection{Research opportunities}

\textit{Opportunity 1:} As indicated above, to accommodate diverse types of workloads, say read-intensive or write-intensive ones, it might be essential to study how to generate diverse workloads and utilize them for pre-fine-tuning in the future.



\noindent
\textit{Opportunity 2:} As mentioned above, it would be necessary to figure out how to reduce the inference overhead of \llm\ without hurting the \ce\ performance before \llm\ can be integrated into real-world DBMS. This could involve enhancing the inference speed of LLMs or developing a more effective deployment strategy beyond the selective estimation approach used in this paper.

\noindent
\textit{Opportunity 3:} As suggested by Section \ref{sec: ablation}, few-shot in-context learning for \ce\ might be sensitive to the selection of few-shot samples. Therefore, a thorough study of appropriate sample selection strategies would be essential.


%% file: conclusion.tex
\section{Conclusion}
\label{sec: concl}
In this paper, we investigate the capability of Large Language Models (\llm) to serve as cardinality estimators, This aims to address the critical challenges that the state-of-the-art ML-based cardinality estimators face, including the poor generalizability to new data or query distribution, inability to
handle complex queries beyond simple SPJ queries, and substantial data preparation overhead. To resolve these issues, we propose preliminary prompting, fine-tuning, and inference strategies for \llm. Additionally, extensive experiments across various settings demonstrate that in comparison to the state-of-the-art, \llm\ exhibit superior in-distribution and out-of-distribution generalizability, enhanced capability to support complex queries, and higher training data efficiency during fine-tuning \llm\ on pre-training datasets. We further evaluate the end-to-end query execution time using the cardinalities estimated by \llm, which suggests that the performance benefits of \llm\ can exceed their inference overhead in some practical settings, thus paving the way toward the integration of \llm\ into the real-world query optimizer. 